\documentclass[12pt,twoside,pointlessnumbers,nofootinbib,smallheadings]{revtex4}

\usepackage[CJKbookmarks]{hyperref}
\usepackage{CJK}
\usepackage{amsmath}
\usepackage{graphics}
\usepackage{slashed}
\usepackage{graphicx}
\usepackage{indentfirst}
\usepackage{amssymb}
\usepackage{cancel}
\usepackage{leftidx}
\usepackage{simplewick}
\usepackage[page,title,titletoc,header]{appendix}

\begin{document}

\begin{CJK*}{GBK}{song}

\title{Spontaneous CP-violation in the Simplest Little Higgs Model and Its Future Collider Tests: the Scalar Sector}

\author{Ying-nan Mao $^{1,}$\footnote{maoyn@ihep.ac.cn}}

\affiliation{
$ ^1$ Center for Future High Energy Physics $\&$ Theoretical Physics Division,
Institute of High Energy Physics, Chinese Academy of Sciences, Beijing 100049, China
}

\begin{abstract}

We proposed the spontaneous CP-violation in the Simplest Little Higgs model. In this model, the pseudoscalar field can acquire
a nonzero vacuum expected value. It leads to a mixing between the two scalars with different CP-charge, which means spontaneous
CP-violation happens. It is also a connection between composite Higgs mechanism and CP-violation. Facing the experimental constraints,
the model is still alive for both scenarios in which the extra scalar appears below or around the electro-weak scale. We also discussed
the future collider tests on CP-violation in the scalar sector through measuring $h_2ZZ$ and $h_1h_2Z'$ vertices (see the definitions
of the particles in the text) which provides new motivations on future $e^+e^-$ and $pp$ colliders. It also shows the importance of
the vector-vector-scalar- and vector-scalar-scalar-type vertices to discover CP-violation effects in the scalar sector.

\end{abstract}

\date{\today}

\maketitle

\newpage

\section{Introduction}
The discovery of a 125 GeV Higgs boson \cite{mass,PDG} by the ATLAS and CMS collaborations \cite{higdisc} in 2012 implies the success of the
standard model (SM) because the measured signal strengths are consistent with those predicted by the SM \cite{stra,strc}. However, the electro-weak
symmetry breaking (EWSB) mechanism is an important topic and researches on physics beyond the SM (BSM) are still necessary and attractive.

For example, to solve the little hierarchy problem, Arkani-Hamed {\em et al.} proposed the Little Higgs (LH) framework \cite{LH} in which the collective symmetry breaking (CSB) mechanism \cite{LH} was used to forbid the quadratic divergences in the Higgs potential at one-loop level. The LH framework contains a lot of models. All of them are special kinds of composite Higgs models \cite{comp} thus each of them must contain a global symmetry which is spontaneously broken at a high scale $f\gg v$ where $v=246~\textrm{GeV}$ is the vacuum expected value (VEV) of the Higgs field. The SM-like Higgs boson is treated as a pseudo-Nambu-Goldstone boson corresponding to one of the broken generators and EWSB is generated dominantly through quantum correction thus the Higgs boson can be naturally light \cite{LH,comp}. Usually the gauge group is also enlarged thus there are extra
gauge bosons with their masses at $\mathcal{O}(f)$ scale. LH models are effective field theories (EFT) below a cut off scale $\Lambda\sim4\pi f$. Below the scale $\Lambda$, a LH model is weakly coupled, but we do not know what would happen above $\Lambda$. Among those models, the simplest Little Higgs (SLH) model \cite{SLH,SLHi,SLH2} has the minimal extended scalar sector in which their are only two scalars. In the SLH
model, a global symmetry $\left[\textrm{SU}(3)\times\textrm{U}(1)\right]^2$ is spontaneously broken to $\left[\textrm{SU}(2)\times\textrm{U}(1)\right]^2$ at scale $f$. The gauge symmetry is enlarged to $\textrm{SU}(3)\times\textrm{U}(1)$ and spontaneously broken to the electro-weak (EW) gauge symmetry $\textrm{SU}(2)_L\times\textrm{U}(1)$ at scale $f$ as well. And at the EW scale $v$, the gauge symmetry is further broken to $\textrm{U}(1)_{\textrm{em}}$ like
what happens in the SM. If CP-violation is absent in the scalar sector, one of the scalars is the SM-like Higgs boson (denoted as $h$), and the other is a pseudoscalar.

CP-violation is another important topic in both SM and BSM physics. In 1964, CP-violation was first discovered through the $K_L\rightarrow\pi\pi$ rare decay process \cite{CPVdisc}. More CP-violation effects have been discovered in K- and B-meson sectors \cite{PDG}. All these measured CP-violation effects can be
successfully explained by the Kobayashi-Maskawa (K-M) mechanism \cite{KM} which was proposed by Kobayashi and Maskawa in 1973. They showed that a nontrivial CP-phase can appear in the quark mixing matrix (named CKM matrix \cite{KM,Cab}) if there exist three generations of fermions. However, the succeed of K-M
mechanism is not the end of CP-violation studies. For example, the observed matter-antimatter asymmetry in the universe \cite{PDG,Plank} requires new sources
of CP-violation because SM itself cannot generate such a large asymmetry \cite{EWBG,EWBG2}. Thus it is attractive to study new CP-violation sources. Till now,
the scalar sector is still an unfamiliar world for us and there may be lots of hidden new physics, including new sources of CP-violation. Thus in this paper,
we focus on extra CP-violation in the scalar sector.

Theoretically, there are already many extensions of the SM which contains new CP-violation sources. For example, if we add more complex scalar singlets or doublets, there may be CP-violation in scalar sector \cite{2HDM,example1,Lee,example2,example3} which can leads to a CP-mixing Higgs boson \footnote{For the 125 GeV Higgs boson, LHC measurements preferred a CP-even one and excluded a CP-odd one at over $3\sigma$ level through the final distribution of $h\rightarrow ZZ^*\rightarrow4\ell$ decay assuming no CP-violation in the Higgs interactions \cite{higCP}. However, a CP-mixing Higgs boson is still allowed since the contribution from pseudoscalar component should be loop suppressed.}. Some of these models may be CP-conserving at the Lagrangian level and CP-violation can arise only from a complex vacuum,
which was called the spontaneous CP-violation mechanism \cite{Lee}. This mechanism was proposed by Lee in 1973 \cite{Lee} as the first kind of two-Higgs-doublet model (2HDM) \cite{2HDM}. Moreover, spontaneous CP-violation mechanism is also a possible solution of the strong-CP problem \cite{strong}, and it may have further connection with lightness of the Higgs boson as well \cite{mao}. Besides these models, spontaneous CP-violation in the scalar sector can also arise from the composite framework. There are already two examples, one is the next-to-minimal composite Higgs model ($\textrm{SO}(6)/\textrm{SO}(5)$, or equivalently $\textrm{SU}(4)/\textrm{Sp}(4)$) \cite{NMC}, and the other is the Littlest Higgs model ($\textrm{SU}(5)/\textrm{SO}(5)$) \cite{CPVLH}. In each model, CP-violation occurs when the pseudoscalar field acquires a nonzero VEV. In this paper, we will propose the possibility of spontaneous CP-violation in the
SLH model through the realization of the same mechanism. This model can also appear as one of the candidates to solve strong-CP problem as mentioned above.
More details on this topic will appear in a forthcoming paper \cite{strongmao}.

Phenomenologically, we can test new CP-violation effects directly or indirectly. The indirect effects may appear in the electric dipole moments (EDM) of
electron and neutron \cite{EDM}, modifications in meson mixing parameters \cite{modify}, or anomalous $ZZZ$ couplings \cite{anoZ}; while the direct effects
may be discovered in $h\tau^+\tau^-$ or $ht\bar{t}$ vertices through measuring the final state distributions \cite{dist}. If another scalar is discovered and
we denote the scalars as $h_{1,2}$ ($h_1$ is the SM-like Higgs boson and $h_2$ is the extra scalar), we can also discover CP-violation in the scalar sector through directly measuring tree-level vector-vector-scalar- ($VVS$-) and vector-scalar-scalar- ($VSS$-) type vertices, such as $h_2VV$ and $Vh_1h_2$ vertices \footnote{Here $V$ denotes a massive gauge boson. For the SM gauge group, $V=W$ or $Z$; while for LH gauge groups, $V$ can also denotes extra heavy gauge bosons.}, according to the CP-properties analysis \cite{mao}. Based on this idea, the author and his collaborators recently proposed a model-independent method to measure the CP-violation effects in the scalar sector through $e^+e^+\rightarrow Z^*\rightarrow Zh_1,Zh_2,h_1h_2$ associated production processes at future $e^+e^-$ colliders \cite{testCPV}. In that research, the product of the three vertices was used as a quantity to measure the magnitude of CP-violation \cite{testCPV,K}. However, in the SLH model, the author and his collaborators recently showed the $Zh_1h_2$ vertex is suppressed by a factor $(v/f)^3$ \cite{Zh1h2} which means it is difficult to test. Thus to test CP-violation in the SLH model, we can turn to extra heavy gauge bosons for help.

As a summary, the model studied in this paper is attractive both theoretically and phenomenologically. This paper is organized as following:
in \autoref{model} we briefly review the CP-conserving SLH model, build the SLH model with spontaneous CP-violation, and obtain the domain interactions;
in \autoref{cons} we consider the constraints on this model, especially in the scalar sector; in \autoref{test} we discuss the tests on CP-violation
effects in this model at future $e^+e^-$ or $pp$ colliders; and in \autoref{conc} we present our conclusions and further discussions. In the appendix
\autoref{form}, we also presented the improved SLH formalism \cite{Zh1h2} which is very helpful for the model building.

\section{Model Construction}
\label{model}
In this section, we first briefly review the CP-conserving SLH model and then construct the spontaneous CP-violation SLH model. We will also derive the
useful vertices in the spontaneous CP-violation SLH model. In both models, we have the same nonlinear realization for Goldstone bosons. We also have the same particle spectra in both models, while in the CP-violation model, the scalars are both CP-mixing states. The CSB mechanism and loop corrections in the Higgs potential are also similar in both models. The difference comes from an extra explicit global $[\textrm{SU}(3)\times\textrm{U}(1)]^2$ breaking term which is absent in the CP-conserving model.
\subsection{A Brief Review of the CP-conserving SLH Model}
\label{CPC}
The SLH model contains two scalar triplets $\Phi_{1,2}$ which transform as $(\mathbf{3},\mathbf{1})$ and $(\mathbf{1},\mathbf{3})$ respectively
under the global $[\textrm{SU}(3)\times\textrm{U}(1)]^2$ transformation \cite{SLH,SLHi,SLH2,SLH3}. At a scale $f\gg v$,
$[\textrm{SU}(3)\times\textrm{U}(1)]^2$ breaks to $[\textrm{SU}(2)\times\textrm{U}(1)]^2$ and ten Nambu-Goldstone bosons are generated, eight of which
should be eaten by massive gauge bosons during spontaneous gauge symmetry breaking $\textrm{SU}(3)\times\textrm{U}(1)\rightarrow
\textrm{SU}(2)_L\times\textrm{U}(1)\rightarrow\textrm{U}(1)_{\textrm{em}}$. Two physical scalars are finally left. The nonlinear realized scalar
triples can be written as \cite{SLH3}
\begin{equation}
\Phi_1=\textrm{e}^{\textrm{i}\Theta'}\textrm{e}^{\textrm{i}t_{\beta}\Theta}\left(\begin{array}{c}\mathbf{0}_{1\times2}\\fc_{\beta}\end{array}\right),\quad\quad
\Phi_2=\textrm{e}^{\textrm{i}\Theta'}\textrm{e}^{-\textrm{i}\Theta/t_{\beta}}\left(\begin{array}{c}\mathbf{0}_{1\times2}\\fs_{\beta}\end{array}\right);
\end{equation}
where $\beta$ is a mixing-angle between the two scalar triplets. The matrix fields $\Theta$ and $\Theta'$ are separately
\begin{equation}
\Theta\equiv\frac{1}{f}\left(\frac{\eta\mathbb{I}_{3\times3}}{\sqrt{2}}+
\left(\begin{array}{cc}\mathbf{0}_{2\times2}&\phi\\ \phi^{\dag}&0\end{array}\right)\right),\quad\textrm{and}\quad
\Theta'\equiv\frac{1}{f}\left(\frac{G'\mathbb{I}_{3\times3}}{\sqrt{2}}+\left(\begin{array}{cc}\mathbf{0}_{2\times2}&\varphi\\ \varphi^{\dag}&0\end{array}\right)\right),
\end{equation}
in which $\phi\equiv\left((v_h+h-\textrm{i}G)/\sqrt{2},G^-\right)^T$ is the usual Higgs doublet and $\varphi\equiv(y^0,x^-)^T$ is another
complex doublet for Goldstones corresponding to heavy gauge bosons following the conventions in \cite{SLH3}.

The covariant derivative term is
\begin{equation}
\label{kin}
\mathcal{L}=\mathop{\sum}_{i=1,2}\left(D^{\mu}\Phi_i\right)^{\dag}\left(D_{\mu}\Phi_i\right)
\end{equation}
where
\begin{equation}
D_{\mu}\equiv\partial_{\mu}-\textrm{i}g\mathbb{G}_{\mu}.
\end{equation}
$g$ is the weak coupling constant and the gauge fields matrix is \cite{SLH,SLHi,SLH3}
\begin{equation}
\mathbb{G}_{\mu}=\frac{A^3_{\mu}}{2}\left(\begin{array}{ccc}1&&\\&-1&\\&&\end{array}\right)+\frac{A^8_{\mu}}{2\sqrt{3}}\left(\begin{array}{ccc}1&&\\&1&\\&&-2\end{array}\right)
+\frac{1}{\sqrt{2}}\left(\begin{array}{ccc}&W^+&Y^0\\W^-&&X^-\\ \bar{Y}^0&X^+&\end{array}\right)_{\mu}+\frac{t_WB_{\mu}}{3\sqrt{1-t^2_W/3}}\mathbb{I}
\end{equation}
where $\theta_W$ is the EW mixing angle \footnote{In this paper, we denote $s_{\alpha}\equiv\sin\alpha$, $c_{\alpha}\equiv\cos\alpha$, and
$t_{\alpha}\equiv\tan\alpha$, for any angle $\alpha$.} and complex fields $Y^0(\bar{Y}^0)\equiv\left(Y^1\pm\textrm{i}Y^2\right)/\sqrt{2}$.
The terms including $\mathbb{G}_{\mu}\mathbb{G}^{\mu}$ in (\ref{kin}) give the masses of gauge bosons. Before EWSB, $v_h=0$; while after EWSB,
$v_h$ is generated through quantum correction. It must be close to $v$ and their difference arises at $\mathcal{O}\left((v/f)^2\right)$ level.
To the leading order of $(v/f)$, we have \cite{SLH3}
\begin{equation}
m_W=\frac{gv}{2}\quad\textrm{and}\quad m_X=m_Y=\frac{gf}{\sqrt{2}}.
\end{equation}
The other three neutral degrees of freedom will mix with each other at leading order of $(v/f)$ through the matrix \cite{SLH3}
\begin{equation}
\left(\begin{array}{c}A\\Z\\Z'\end{array}\right)=\left(\begin{array}{ccc}-s_W&s_Xc_W&c_Xc_W\\c_W&s_Xs_W&c_Xs_W\\0&c_X&-s_X\end{array}\right)
\left(\begin{array}{c}A^3\\A^8\\B\end{array}\right)_{\mu}
\end{equation}
where $\theta_X\equiv\arcsin\left(t_W/\sqrt{3}\right)$. The corresponding masses at leading order of $(v/f)$ are then
\begin{equation}
m_A=0,\quad m_Z=\frac{gv}{2c_W},\quad\textrm{and}\quad m_{Z'}=\sqrt{\frac{2}{3-t^2_W}}gf.
\end{equation}
The massless gauge boson is photon. If we go beyond leading order of $(v/f)$, the gauge bosons will have further mixing
with each other. For example, in charged sector, $W^{\pm}$ and $X^{\pm}$ will mix with each other at $\mathcal{O}\left((v/f)^3\right)$ level,
and $W(X)^{\pm}$ will acquire their relative mass corrections at $\mathcal{O} \left((v/f)^2\right)$ level.
While in neutral sector, the off-diagonal elements of the mass matrix $\mathbb{M}^2_V$ in the basis $(Z,Z',Y^2)$ are nonzero. Using an orthogonal
matrix $\mathbb{R}$, it can be diagonalized as $\left(\mathbb{R}\mathbb{M}^2_V\mathbb{R}^T\right)_{pq}=m_p\delta_{pq}$ where $m_p$ are the gauge
bosons' masses. The neutral gauge bosons acquire their mass corrections as
\begin{equation}
\delta m^2_Z=-\delta m^2_{Z'}=\frac{g^2v^2c^2_{2W}}{32c^6_W}\left(\frac{v}{f}\right)^2,\quad\textrm{and}\quad
\delta m^2_{Y^2}=0.
\end{equation}
We denote the corresponding mass eigenstates as $\tilde{Z}$, $\tilde{Z}'$, and $\tilde{Y}^2$. Their mixing angles
(which are also approximately the rotation matrix elements)
\begin{equation}
\mathbb{R}_{Z'Z}=\frac{\sqrt{3}c_{2W}c_X}{8c^3_W}\left(\frac{v}{f}\right)^2,\quad
\mathbb{R}_{Y^2Z}=\frac{\sqrt{2}}{3t_{2\beta}c_W}\left(\frac{v}{f}\right)^3,\quad\textrm{and}\quad
\mathbb{R}_{Y^2Z'}=\frac{2c_X}{\sqrt{6}t_{2\beta}}\left(\frac{v}{f}\right)^3;
\end{equation}
to the leading order of $(v/f)$. $A$ and $Y^1$ do not participate further mixing.

The six neutral scalar degrees of freedom can be divided into CP-even ($h$ and $y^1$) and CP-odd ($\eta$, $G$, $G'$, and $y^2$) parts where
$y^0(\bar{y}^0)\equiv\left(y^1\pm\textrm{i}y^2\right)/\sqrt{2}$. A straightforward calculation showed that after EWSB, the kinetic terms
can be written as
\begin{equation}
\mathcal{L}_{\textrm{kin}}=\frac{1}{2}\left(\partial^{\mu}h\partial_{\mu}h+\partial^{\mu}y^1\partial_{\mu}y^1+\mathbb{K}_{ij}\partial^{\mu}G_i\partial_{\mu}G_j\right)
\end{equation}
with $G_i$ runs over the four CP-odd scalar degrees of freedom and $\mathbb{K}_{ij}\neq\delta_{ij}$ means the CP-odd part is not
canonically-normalized \footnote{Details on the improved formalism to treat this case can be found in the appendix \autoref{form}
and \cite{Zh1h2}.}. To find out the canonically-normalized basis, we should consider the gauge fixing term together. The two-point
transitions between gauge bosons and scalars arise from the cross terms of $\partial_{\mu}\Phi_i$ and $\mathbb{G}_{\mu}\Phi_i$.
These transitions can be parameterized as $V_p^{\mu}\mathbb{F}_{pi}\partial_{\mu}G_i$ and their contributions should be canceled
by $\left(\partial_{\mu}V_p^{\mu}\right)\mathbb{F}_{pi}G_i$ from the gauge fixing term. It can be checked straightforwardly that
(see appendix \autoref{form} or \cite{Zh1h2} for more details) a new basis
\begin{equation}
\label{basis}
\left(\tilde{\eta},\tilde{G}_p\right)=
\left(\frac{\eta}{\sqrt{\left(\mathbb{K}^{-1}\right)_{11}}},\frac{\left(\mathbb{RF}\right)_{pi}}{m_p}G_i\right)
\end{equation}
is canonically-normalized. $\tilde{G}_p$ is just the corresponding Goldstone of $\tilde{V}_p$.

In the fermion sector, each left-handed doublet must be extended to a triplet thus there must be additional heavy fermions.
In lepton sector, a heavy neutrino $N_i$ should be added for each generation. While in the quark sector, choosing the
``anomaly-free embedding" \cite{af}, $T$ with $Q=2/3$ is added as the parter of $t$, $D$ and $S$ with $Q=-1/3$ are added as the
parters of $d$ and $s$ separately. The Yukawa interactions are then \cite{SLH,SLHi,SLH2,SLH3}
\begin{eqnarray}
\label{Yukawa}
\mathcal{L}_y&=&\textrm{i}\lambda^j_N\bar{N}_{R,j}\Phi^{\dag}_2L_j
-\frac{\textrm{i}\lambda_{\ell}^{jk}}{\Lambda}\bar{\ell}_{R,j}\det\left(\Phi_1,\Phi_2,L_k\right)\nonumber\\
&&+\textrm{i}\left(\lambda^a_t\bar{u}_{R,3}^a\Phi_1^{\dag}+\lambda^b_t\bar{u}_{R,3}^b\Phi_2^{\dag}\right)Q_3
-\textrm{i}\frac{\lambda_{b,j}}{\Lambda}\bar{d}_{R,j}\det\left(\Phi_1,\Phi_2,Q_3\right)\nonumber\\
&&+\textrm{i}\left(\lambda_{d,n}^a\bar{d}_{R,n}^a\Phi_1^T+\lambda_{d,n}^b\bar{d}_{R,n}^b\Phi_2^T\right)Q_n
-\textrm{i}\frac{\lambda_u^{jk}}{\Lambda}\bar{u}_{R,j}\det\left(\Phi_1^*,\Phi_2^*,Q_k\right);
\end{eqnarray}
where the left-handed triplets are \cite{SLH3}
\begin{eqnarray}
&L_i=\left(\nu_L,\ell_L,\textrm{i}N_L\right)^T_i,\quad Q_1=\left(d_L,-u_L,\textrm{i}D_L\right)^T,&\nonumber\\
&Q_2=\left(s_L,-c_L,\textrm{i}S_L\right)^T,\quad Q_3=\left(t_L,b_L,\textrm{i}T_L\right)^T.&
\end{eqnarray}
The first line is for leptons where $\ell_{R,j}$ runs over $(e,\mu,\tau)_R$; the second line is for the third generation of quarks where
$d_{R,j}$ runs over $(d,s,b,D,S)_R$, and the last line is for the first two generations of quarks where $u_{R,j}$ runs over $(u,c,t,T)_R$.
$\Lambda\sim4\pi f$ is a cut-off scale. A right-handed quark with index $a$ or $b$ must be a mixing state between an additional quark and
its SM partner, for example, $u_{R,3}^{a,b}$ are mixing states between $t_R$ and $T_R$. To the leading order of $(v/f)$, The heavy fermions'
masses are \cite{SLHi,SLH3}
\begin{equation}
\label{mheavy}
m_N^j=\lambda_N^jfs_{\beta},\quad m_Q=\sqrt{\left|\lambda_q^ac_{\beta}\right|^2+\left|\lambda_q^bs_{\beta}\right|^2}f,
\end{equation}
for $Q=T,D,S$ and $q=t,d(d_1),s(d_2)$. To the leading order, the corresponding partners in SM sector have the masses
\begin{equation}
m_{\nu}^j=0,\quad m_q=\frac{v}{\sqrt{2}}\frac{\left|\lambda_q^a\lambda_q^b\right|}{\sqrt{\left|\lambda_q^ac_{\beta}\right|^2+\left|\lambda_q^bs_{\beta}\right|^2}}=\frac{\lambda_qv}{\sqrt{2}}.
\end{equation}
CSB mechanism keeps all neutrinos massless \footnote{In the first term of (\ref{Yukawa}), we can also use $\Phi_1$
instead of $\Phi_2$, but we cannot have both terms together if we assume massless neutrinos. If we perform this replacement, $m_N^j$ in (\ref{mheavy})
should also be changed to $\lambda_N^jfc_{\beta}$.}. Other fermions require their masses (similarly, to the leading order)
\begin{equation}
m_{\ell}^j=\frac{v}{4\sqrt{2}\pi}y_{\ell}^j,\quad m_b=\frac{v}{4\sqrt{2}\pi}\lambda_{b,3},\quad m_{u,c}=\frac{v}{4\sqrt{2}\pi}y_{u,c},
\end{equation}
in which $y^j_{\ell}$ are eigenvalues of matrix $\lambda_{\ell}^{jk}$ and $y_{u,c}$ are eigenvalues of matrix $\lambda_u^{jk}$. To this step, we
ignored small mixing between $q$ and $Q$. Consider this kind of mixing $\Delta_{qQ}$, a mass correction $\delta m_q/m_q\sim\mathcal{O}
\left(\Delta_{qQ}^2/m^2_Q\right)$ is generated.

Last, let's turn to the scalar potential. In the discussions above, we assume the Higgs doublet acquire a correct VEV to derive the particle spectra
everywhere. However, at tree-level, $\left|\Phi_1^{\dag}\Phi_2\right|^2$ term is forbidden due to the CSB mechanism. The Higgs potential can be
generated through Coleman-Weinberg mechanism \cite{CW} at loop-level as
\begin{equation}
\label{CW}
\delta V_h=-\delta m^2\left(h^{\dag}h\right)+\delta\lambda\left(h^{\dag}h\right)^2.
\end{equation}
The CSB mechanism forbids quadratic divergence in (\ref{CW}) thus \cite{SLH,SLHi,SLH2}
\begin{eqnarray}
\label{dm}
\left(\delta m^2\right)_{\textrm{1-loop}}&=&\frac{3}{8\pi^2}\left(\lambda^2_tm^2_T\ln\frac{\Lambda^2}{m^2_T}
-\frac{g^2m^2_X}{4}\ln\frac{\Lambda^2}{m^2_X}-\frac{g^2m^2_{Z'}\left(1+t^2_W\right)}{8}\ln\frac{\Lambda^2}{m^2_{Z'}}\right);\\
\label{dla}
\left(\delta\lambda\right)_{\textrm{1-loop}}&=&\frac{\left(\delta m^2\right)_{\textrm{1-loop}}}{3f^2s^2_{\beta}c^2_{\beta}}+\frac{3}{16\pi^2}
\Bigg(\lambda^4_t\left(\ln\frac{m^2_T}{m^2_t}-\frac{1}{2}\right)\nonumber\\
&&-\frac{g^4}{8}\left(\ln\frac{m^2_X}{m^2_W}-\frac{1}{2}\right)-\frac{g^4\left(1+t^2_W\right)^2}{16}\left(\ln\frac{m^2_{Z'}}{m^2_Z}-\frac{1}{2}\right)\Bigg).
\end{eqnarray}
Here $\Lambda\sim4\pi f$ is a cut-off scale and $\lambda_t\equiv\sqrt{2}m_t/v$, which means the contributions from the first and second generations
of fermions are ignorable. When $m_T$ is heavy enough, EWSB can be generated through these loop corrections.

Now the pseudoscalar $\eta$ is still massless due to an accidental global $\textrm{U}(1)$ symmetry. Adding a term
\begin{equation}
\label{add}
\delta V=-\mu^2\Phi_1^{\dag}\Phi_2+\textrm{H.c.}
\end{equation}
in the potential \footnote{This term breaks the CSB mechanism explicitly which means a quadratic divergence in the Higgs potential
can be generated at one-loop level. Thus numerically $\mu$ should be very small comparing with $f$. In the convention of this paper
(which is the same as that in \cite{SLH3}), the degrees of freedom in $\Theta'$ cancels with each other thus $\eta$ dose not acquire
additional mixing with $y^2$.}, $\eta$ acquires its mass \cite{SLH2}
\begin{equation}
m^2_{\eta}=\frac{\mu^2}{s_{\beta}c_{\beta}}\cos\left(\frac{v}{\sqrt{2}fs_{\beta}c_{\beta}}\right)\approx\frac{\mu^2}{s_{\beta}c_{\beta}}
\end{equation}
and the Higgs potential acquires another correction \cite{SLH2}
\begin{eqnarray}
\left(\delta V_h\right)_{\mu}&=&-\left(\delta m^2\right)_{\mu}\left(h^{\dag}h\right)+\left(\delta\lambda\right)_{\mu}\left(h^{\dag}h\right)^2\nonumber\\
\label{lammu}
&=&m^2_{\eta}\left(h^{\dag}h\right)-\frac{m^2_{\eta}}{12f^2s^2_{\beta}c^2_{\beta}}\left(h^{\dag}h\right)^2.
\end{eqnarray}
Two-loop contributions to $\delta m^2$ can be absorbed into the possible contributions from
unknown physics at the cut-off scale $\Lambda$ \cite{LH,SLHa} which can be parameterized as $\left(\delta m^2\right)_{\textrm{2-loop}}=-cf^2$.
We can roughly estimate $|c|\sim\mathcal{O}(10^{-2})$.

\subsection{Spontaneous CP-violation in the SLH Model}
\label{CPV}
In (\ref{add}), $\mu$-term provides the $\eta$ mass. In general, $\mu^2$ can be complex, but its argument can always be absorbed into the
shift of $\eta$ (which is equivalent to a rotation of $\Phi_i$). Besides this, $\eta$ cannot acquire a nonzero VEV, thus there is no CP-violation
in the scalar potential. Comparing with the CP-conserving case in \autoref{CPC}, we can add another term and (\ref{add}) becomes
\begin{equation}
\label{add2}
\delta V=-\mu^2\Phi_1^{\dag}\Phi_2+\epsilon\left(\Phi_1^{\dag}\Phi_2\right)^2+\textrm{H.c.}
\end{equation}
Here $\epsilon$ is also required to be small (for example, $\epsilon\lesssim\mathcal{O}\left((v/f)^2\right)$ thus the CSB mechanism
is not significantly broken). In general, $\mu^2$ and $\epsilon$ can be complex, but we can shift $\eta$ to make at least one of them
real. If we choose $\mu^2$ real, when $\epsilon$ is still complex, CP-symmetry would be explicitly broken in the scalar sector. However,
if both $\mu^2$ and $\epsilon$ are real, $\eta$ is also possible to acquire a nonzero VEV which means spontaneous CP-violation happens.
In this paper, we focus on the spontaneous CP-violation case.

According to (\ref{add2}), denote $\alpha\equiv v_h/(\sqrt{2}fs_{\beta}c_{\beta})$, we have
\begin{equation}
V_{\eta}=-\mu^2f^2s_{\beta}c_{\beta}c_{\alpha}\cos\left(\frac{\eta}{\sqrt{2}fs_{\beta}c_{\beta}}\right)+
\epsilon f^4s^2_{\beta}c^2_{\beta}c^2_{\alpha}\cos\left(\frac{\sqrt{2}\eta}{fs_{\beta}c_{\beta}}\right).
\end{equation}
Minimize this potential, we found that when
\begin{equation}
\mu^2<4\epsilon f^2\left|s_{\beta}c_{\beta}c_{\alpha}\right|,
\end{equation}
$\langle\eta\rangle=0$ becomes unstable thus $\eta$ would acquire a nonzero VEV
\begin{equation}
v_{\eta}\equiv\langle\eta\rangle=\pm\sqrt{2}fs_{\beta}c_{\beta}\arccos\left(\frac{\mu^2}{4\epsilon f^2s_{\beta}c_{\beta}c_{\alpha}}\right),
\end{equation}
which means spontaneous CP-violation is possible. For simplify, we choose ``+" in the equation above from now on. We denote $\xi\equiv v_{\eta}/(\sqrt{2}fs_{\beta}c_{\beta})$, and the scalar mass term is
\begin{equation}
\mathcal{L}_{\textrm{m}}=-\frac{1}{2}\left(h,\eta\right)\left(\begin{array}{cc}M^2&\epsilon f^2s_{2\alpha}s_{2\xi}\\ \epsilon f^2s_{2\alpha}s_{2\xi}&4\epsilon f^2c^2_{\alpha}s_{\xi}^2\end{array}\right)\left(\begin{array}{c}h\\ \eta\end{array}\right).
\end{equation}
Here $M$ should be close to $125~\textrm{GeV}$ and it includes all the quantum correction effects from (\ref{dm}) and (\ref{dla}) \footnote{These quantum
corrections are not affected by the CP properties of the scalar sector which means (\ref{dm}) and (\ref{dla}) derived in the CP-conserving model
can be simply transported into the CP-violation case}. Nonzero off-diagonal elements means the mass eigenstates cannot be CP eigenstates. Define
the mass eigenstates (in which $h_1$ is SM-like)
\begin{equation}
\left(\begin{array}{c}h_1\\h_2\end{array}\right)\equiv\left(\begin{array}{cc}c_{\theta}&-s_{\theta}\\s_{\theta}&c_{\theta}\end{array}\right)
\left(\begin{array}{c}h\\ \eta\end{array}\right),
\end{equation}
we have the mixing angle
\begin{equation}
\theta=\frac{1}{2}\arctan\left(\frac{2\epsilon f^2s_{2\alpha}s_{2\xi}}{M^2-4\epsilon f^2c^2_{\alpha}s^2_{\xi}}\right)
\end{equation}
and scalar masses
\begin{equation}
\label{mass}
m_{1,2}=\sqrt{\frac{M^2+4\epsilon f^2c^2_{\alpha}s^2_{\xi}}{2}\pm\left(\frac{M^2-4\epsilon f^2c^2_{\alpha}s^2_{\xi}}{2}c_{2\theta}
+\epsilon f^2s_{2\alpha}s_{2\xi}s_{2\theta}\right)}.
\end{equation}
We can see that only when both $\mu^2$ and $\epsilon$ are nonzero, CP-violation can occur, which means in this model, CP-symmetry
is also collectively broken \footnote{The case $\epsilon$ absents was already discussed above. The case $\mu^2$ absents allows a
nonzero $v_{\eta}$, but $\xi=\pi/2$ that the off-diagonal elements in (\ref{mass}) are still zero. A shift of $\eta$ (rotation of $\Phi$)
can remove this $\xi$ hence it is trivial. A nontrivial $\xi$ requires nontrivial $\mu^2$ and $\epsilon$.}.

For the Yukawa couplings, we can also choose all the couplings real thus there is no explicit CP-violation. Complex CKM matrix can arise
from the mixing between a SM quark and an extra quark, which is the same mechanism as that in \cite{example1}.

\subsection{Some Useful Interactions in this Model}
In the CP-violation SLH model, mixing between $h$ and $\eta$ can modify some of the vertices in the CP-conserving model. The $hVV$
couplings can be parameterized as
\begin{equation}
\mathcal{L}_{hVV}=\frac{g^2v}{2}\mathop{\sum}_{V}\left(\left(\tilde{c}_{1,V}h_1+\tilde{c}_{2,V}h_2\right)\tilde{V}\tilde{V}^*\right)
\end{equation}
where $\tilde{V}$ denote the mass eigenstates. For real vector fields, $\tilde{V}^*=\tilde{V}$. To the leading order of $(v/f)$, we have
\begin{eqnarray}
\tilde{c}_{1,W}=-\tilde{c}_{1,X}=c_{\theta},&\quad&\tilde{c}_{2,W}=-\tilde{c}_{2,X}=s_{\theta},\\
\tilde{c}_{1,Z}=-\tilde{c}_{1,Z'}=\frac{c_{\theta}}{2c^2_W},&\quad&\tilde{c}_{2,Z}=-\tilde{c}_{2,Z'}=\frac{s_{\theta}}{2c^2_W},
\end{eqnarray}
while $c_{i,Y}$ remains zero to all order of $(v/f)$.

For the antisymmetric type $Vh\eta$ couplings \footnote{We don't consider the symmetric type couplings $(h_1\partial^{\mu}h_2+h_2\partial^{\mu}h_1)$
here because they cannot contribute anything in the processes with on-shell gauge boson(s).}, we parameterize it as
\begin{equation}
\mathcal{L}_{Vh_1h_2}=\frac{g}{2}\left(h_1\partial^{\mu}h_2-h_2\partial^{\mu}h_1\right)\left(\tilde{c}^{as}_{Zh_1h_2}\tilde{Z}_{\mu}
+\tilde{c}^{as}_{Z'h_1h_2}\tilde{Z}'_{\mu}+\tilde{c}^{as}_{Yh_1h_2}\tilde{Y}^2_{\mu}\right).
\end{equation}
The results to the leading order of $(v/f)$ are
\begin{equation}
\label{Vh1h2}
\tilde{c}^{as}_{Zh_1h_2}=\frac{1}{2\sqrt{2}c^3_Wt_{2\beta}}\left(\frac{v}{f}\right)^3,\quad\tilde{c}^{as}_{Z'h_1h_2}=
\frac{2\sqrt{2}}{\sqrt{3-t^2_W}t_{2\beta}}\left(\frac{v}{f}\right),\quad\tilde{c}^{as}_{Yh_1h_2}=-1;
\end{equation}
which are the same as the CP-conserving case, since $h_1\partial^{\mu}h_2-h_2\partial^{\mu}h_1=h\partial^{\mu}\eta-\eta\partial^{\mu}h$.

The scalar trilinear interactions should be
\begin{equation}
\mathcal{L}_S=-\frac{1}{2}\lambda_{122}fh_1h^2_2-\frac{1}{2}\lambda_{211}fh_2h^2_1;
\end{equation}
where to the leading order of $(v/f)$, the dimensionless coefficients
\begin{eqnarray}
\lambda_{122}&=&c_{\theta}\left(1-3s^2_{\theta}\right)\frac{\sqrt{2}\epsilon s_{2\alpha}\left(3c_{2\xi}-1\right)}{s_{2\beta}}
+s_{\theta}\left(2-3s^2_{\theta}\right)\frac{\sqrt{2}\epsilon s_{2\xi}\left(3c_{2\alpha}-1\right)}{s_{2\beta}}\nonumber\\
&&-6c^2_{\theta}s_{\theta}\frac{\sqrt{2}\epsilon c^2_{\alpha}s_{2\xi}}{s_{2\beta}}+6c_{\theta}s^2_{\theta}\frac{\lambda v}{f},\\
\lambda_{211}&=&c_{\theta}\left(1-3s^2_{\theta}\right)\frac{\sqrt{2}\epsilon s_{2\xi}\left(3c_{2\alpha}-1\right)}{s_{2\beta}}
-s_{\theta}\left(2-3s^2_{\theta}\right)\frac{\sqrt{2}\epsilon s_{2\alpha}\left(3c_{2\xi}-1\right)}{s_{2\beta}}\nonumber\\
&&+6c^2_{\theta}s_{\theta}\frac{\lambda v}{f}+6c_{\theta}s^2_{\theta}\frac{\sqrt{2}\epsilon c^2_{\alpha}s_{2\xi}}{s_{2\beta}}.
\end{eqnarray}
$\lambda$ in the equations is the Higgs self-coupling constant.

The Yukawa couplings for SM leptons and quarks $f=\ell,q$ can be parameterized as
\begin{equation}
\mathcal{L}_{\textrm{y}}=-\mathop{\sum}_{f}\frac{m_f}{v}\left(\left(c_{1,f}h_1+c_{2,f}h_2\right)\bar{f}_Lf_R\right)+\textrm{H.c.}
\end{equation}
For $f=u,c,b,\nu,\ell$, the pseudoscalar degree of freedom dose not couple to these fermions, thus we have
\begin{equation}
c_{1,f}=c_{\theta}\quad\textrm{and}\quad c_{2,f}=s_{\theta};
\end{equation}
while for $q=d,s,t$, the coupling coefficients
\begin{equation}
c_{1,q}=c_{\theta}+\textrm{i}\delta_qs_{\theta}\frac{v}{f}\frac{c_{2\beta}+c_{2\theta_R}}{\sqrt{2}s_{2\beta}}\quad\textrm{and}\quad
c_{2,q}=s_{\theta}-\textrm{i}\delta_qc_{\theta}\frac{v}{f}\frac{c_{2\beta}+c_{2\theta_R}}{\sqrt{2}s_{2\beta}}.
\end{equation}
Here $\delta_q=-1$ for the third generation ($q=t$) and $\delta_q=+1$ for the first two generations ($q=d,s$). The imaginary parts
are generated by the left-handed mixing between light and heavy quarks. $\theta_R=
\arctan\left(t^{-1}_{\beta}\lambda_1/\lambda_2\right)$ at the leading order of $v/f$ is the right-handed mixing angle.
Here we don't consider the possible flavor changing couplings. The Yukawa couplings including a heavy quark should be
\begin{eqnarray}
\mathcal{L}_{\textrm{Y}}&=&-\mathop{\sum}_Q\frac{m_Q}{f}\Big(\left(c_{1,Q}h_1+c_{2,Q}h_2\right)\bar{Q}_LQ_R\nonumber\\
&&+\bar{q}\left(\left(c_{1L,q}h_1+c_{2L,q}h_2\right)P_L+\left(c_{1R,q}h_1+c_{2R,q}h_2\right)P_R\right)Q+\textrm{H.c.}\Big)
\end{eqnarray}
where $P_{L/R}=(1\mp\gamma^5)/2$. The coefficients
\begin{equation}
c_{1,Q}=-c_{\theta}\frac{v}{2f}\left(\frac{s_{2\theta_R}}{s_{2\beta}}\right)^2
+\textrm{i}\delta_Qs_{\theta}\frac{c_{2\beta}+c_{2\theta_R}}{\sqrt{2}s_{2\beta}},\quad
c_{2,Q}=s_{\theta}\frac{v}{2f}\left(\frac{s_{2\theta_R}}{s_{2\beta}}\right)^2
-\textrm{i}\delta_Qc_{\theta}\frac{c_{2\beta}+c_{2\theta_R}}{\sqrt{2}s_{2\beta}}.
\end{equation}
Here $\delta_Q=+1$ for the third generation ($Q=T$) and $\delta_Q=-1$ for the first two generations ($Q=D,S$), which different with those
for SM fermions. $s_{2\theta_R}\propto\delta_Qm_q/m_Q$ thus for the first two generations, we have $s_{2\theta_R}\ll1$. The other four coefficients
including both light and heavy quarks are
\begin{eqnarray}
\label{c1}
c_{1L,q}&=&c_{\theta}\frac{v}{2f}\frac{(c_{2\beta}-c_{2\theta_R})s_{2\theta_R}}{s^2_{2\beta}}
+\textrm{i}\delta_Qs_{\theta}\frac{s_{2\theta_R}}{\sqrt{2}s_{2\beta}}=
-\delta_Qc_{\theta}\frac{m_q}{\sqrt{2}m_Q}\frac{c_{2\beta}-c_{2\theta_R}}{s_{2\beta}}-\textrm{i}s_{\theta}\frac{m_qf}{m_Qv},\\
c_{2L,q}&=&s_{\theta}\frac{v}{2f}\frac{(c_{2\beta}-c_{2\theta_R})s_{2\theta_R}}{s^2_{2\beta}}
-\textrm{i}\delta_Qc_{\theta}\frac{s_{2\theta_R}}{\sqrt{2}s_{2\beta}}=
-\delta_Qs_{\theta}\frac{m_q}{\sqrt{2}m_Q}\frac{c_{2\beta}-c_{2\theta_R}}{s_{2\beta}}+\textrm{i}c_{\theta}\frac{m_qf}{m_Qv};\\
c_{1R,q}&=&\delta_Qc_{\theta}\frac{c_{2\beta}+c_{2\theta}}{\sqrt{2}s_{2\beta}}
-\textrm{i}s_{\theta}\frac{v}{2f}\left(\left(\frac{c_{2\beta}+c_{2\theta_R}}{s_{2\beta}}\right)^2-1\right),\\
\label{c2}
c_{2R,q}&=&\delta_Qs_{\theta}\frac{c_{2\beta}+c_{2\theta}}{\sqrt{2}s_{2\beta}}
+\textrm{i}c_{\theta}\frac{v}{2f}\left(\left(\frac{c_{2\beta}+c_{2\theta_R}}{s_{2\beta}}\right)^2-1\right).
\end{eqnarray}
In the calculation of $c_{iR,q}$, the improved formalism affects on their imaginary parts since the $\eta$ component in $G$ cannot
be ignored due to the improved SLH formalism \cite{Zh1h2}. For the third generation, $m_t/m_T\sim\mathcal{O}(v/f)$, thus $c_{iL,q}$
can reach $\mathcal{O}(1)$. But for the first two generations, $m_q/m_Q\ll v/f$ means $c_{iL,q}\ll1$.

\section{Recent Constraints on the Model}
\label{cons}
As a BSM model, SLH always face many direct and indirect constraints, such as collider searches for new particles predicted by the
model and EW precision tests. The scalar sector contains an extra scalar $h_2$, whose properties are quite different from the SM-like scalar.
If it is light enough ($m_2<m_1/2$), it should also face the $h_1$ cascade decay constraint. As a model with new CP-violation source, we should
also discuss the EDM constraints \cite{EDM}. In this paper, we don't discuss more details about quark flavor physics.
\subsection{Direct and Indirect Constraints on $f$}
In the SLH model, the modifications on $S$ and $T$ parameters are sensitive to the new scale $f$. Thus before LHC Run II, the $S$
and $T$ parameter constraint \cite{obl1,obl2,obl3} on $f$ used to be the strictest one. $f\gtrsim(4-7)~\textrm{TeV}$ at $95\%$ C.L.
when $t_{\beta}\sim(1-10)$ \cite{SLH4,SLH5}. In the SLH with spontaneous CP-violation, this constraint is similar, because the $S$ and $T$
parameters are note sensitive to $m_2$ and $c_{2,W/Z}$ when $c_{2,W/Z}\ll1$.

However, since LHC Run II began, the lower limits on exotic particles increase quickly hence the corresponding new physics scales are
pushed higher. In the SLH model, $\tilde{X}^{\pm}$ and $\tilde{Y}^0(\tilde{\bar{Y}}^0)$ gauge bosons couple to SM fermions with a
suppression factor $v/f$, thus they are difficult to be produced at LHC. However, couplings between $\tilde{Z}'$ and SM fermions have
the same order with those in SM \footnote{These couplings are the same in CP-conserving and CP-violation models.},
thus $\tilde{Z}'$ searches at LHC can provide a direct constraint on $f$. Recently, using $36.1~\textrm{fb}^{-1}$ luminosity at $\sqrt{s}=
13~\textrm{TeV}$, ATLAS collaboration set a new constraint $m_{Z'}\gtrsim4.5~\textrm{TeV}$ at $95\%$ C.L. \cite{Zprime} for the
sequential standard model (SSM) \cite{SSM} in which $Z'$ couples to SM fermions with the strengths in the SM.

In the SLH model with ``anomaly free embedding", the gauge couplings for fermions are fixed, which can be found in \cite{SLHi} and \cite{SLH3}.
The signal strength are then \cite{PDG,SLH2,SLH3,SSM}
\begin{equation}
\mu\equiv\frac{\left(\sigma_{Z'}\textrm{Br}_{Z'\rightarrow\ell^+\ell^-}\right)_{\textrm{SLH}}}
{\left(\sigma_{Z'}\textrm{Br}_{Z'\rightarrow\ell^+\ell^-}\right)_{\textrm{SSM}}}=0.36\frac{\kappa_{d/u}+1.14}{\kappa_{d/u}+0.78}\approx0.49,
\end{equation}
in which
\begin{equation}
\kappa_{d/u}\equiv\frac{\int dx_1dx_2f_d(x_1)f_{\bar{d}}(x_2)\delta(x_1x_2-m^2_{Z'}/s)}{\int dx_1dx_2f_u(x_1)f_{\bar{u}}(x_2)\delta(x_1x_2-m^2_{Z'}/s)}
\sim(0.2-0.25),
\end{equation}
for $m_{Z'}=(4-4.5)~\textrm{TeV}$, using the MSTW2008 PDF \cite{MSTW}. Comparing with the results shown in \cite{Zprime}
and assuming $m_{T,D,S,N_i}>m_{Z'}/2$, it can be roughly estimated that $f\gtrsim7.5~\textrm{TeV}$ at $95\%$ C.L \footnote{Recently, Dercks {\em et al.}
reported new lower limit $f\gtrsim1.3~\textrm{TeV}$ for littlest Higgs model with T-parity \cite{TParity}, which is quite lower than the limit in SLH
model. That is because in the T-parity model, extra $Z'$ boson is T-odd thus it cannot have sizable coupling with SM fermion pairs. Thus in that model,
direct searches on $Z'$ cannot lead to strict constraint on the scale $f$}. Comparing with the indirect constraints discussed above, we can see that the $Z'$ direct searching experiments can provide the strictest constraint
on $f$ in the SLH model for most $\beta$ region.

\subsection{Constraints on the Properties of Extra Scalar $h_2$}
$h_2$ couples to SM particles dominantly through its $h$ component, since the couplings between $\eta$ component and SM sector
are highly suppressed by the high scale $f$. Experimentally, for a light $h_2$, it mainly face the direct searches through
$e^+e^-\rightarrow Zh_2$ at LEP; while for a heavy $h_2$, it mainly faces the direct searches through $gg\rightarrow h_2
\rightarrow W^+W^-/ZZ$ at LHC. Both production cross sections a suppressed by a factor $s^2_{\theta}$. When $m_2<m_1/2$, it
should also face the $h_1\rightarrow2h_2$ rare decay constraint. Theoretically, the allowed parameter region also depend on
the details of EWSB.

\begin{figure}[h]
\caption{The $h_1\rightarrow2h_2$ decay branching ratio distribution in $s_{\theta}$-$m_2$ plane with $f=8~\textrm{TeV}$. From left to right,
we choose $t_{\beta}=1,3,6$, respectively.}
\label{EXO}
\includegraphics[scale=0.43]{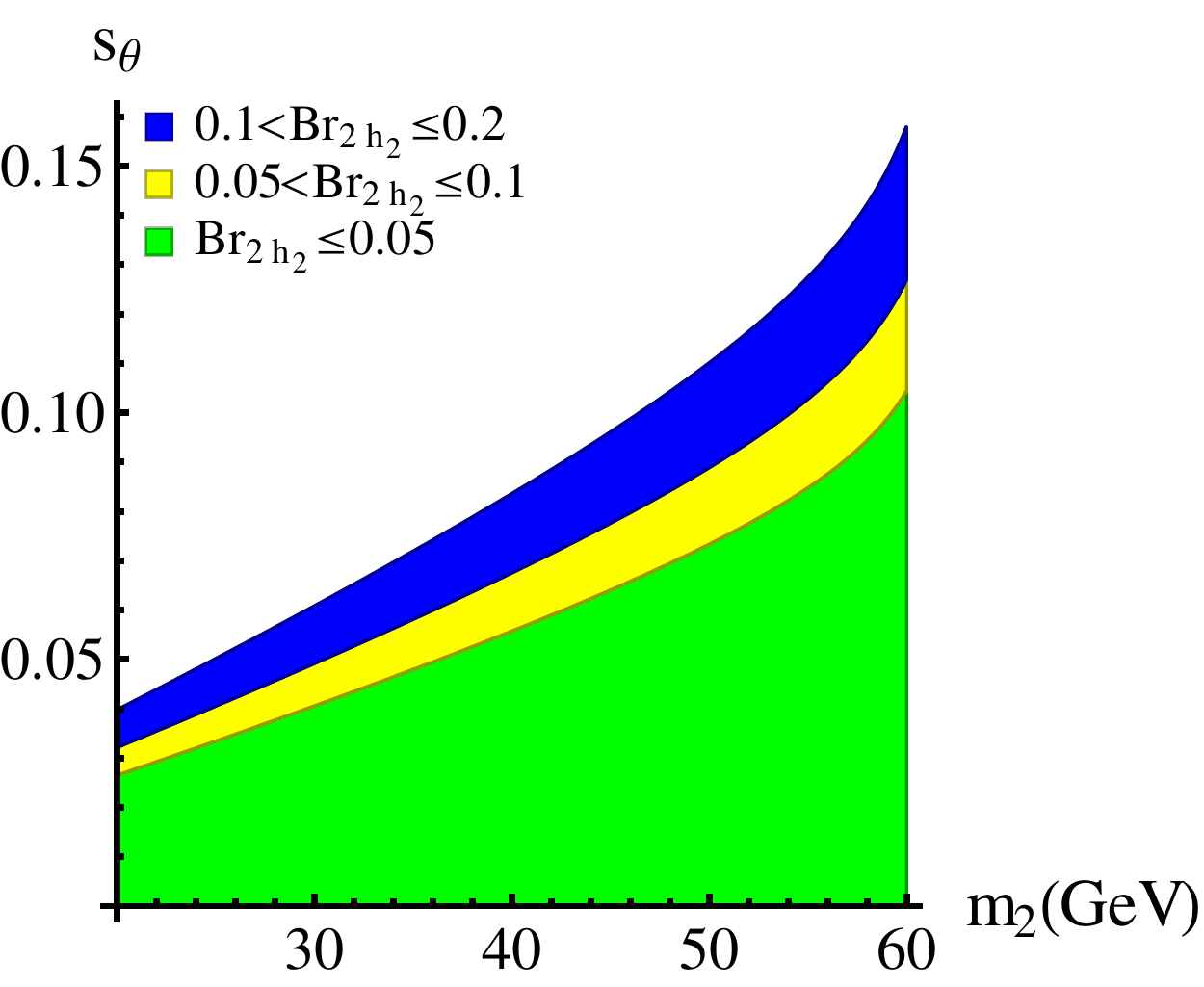}\includegraphics[scale=0.43]{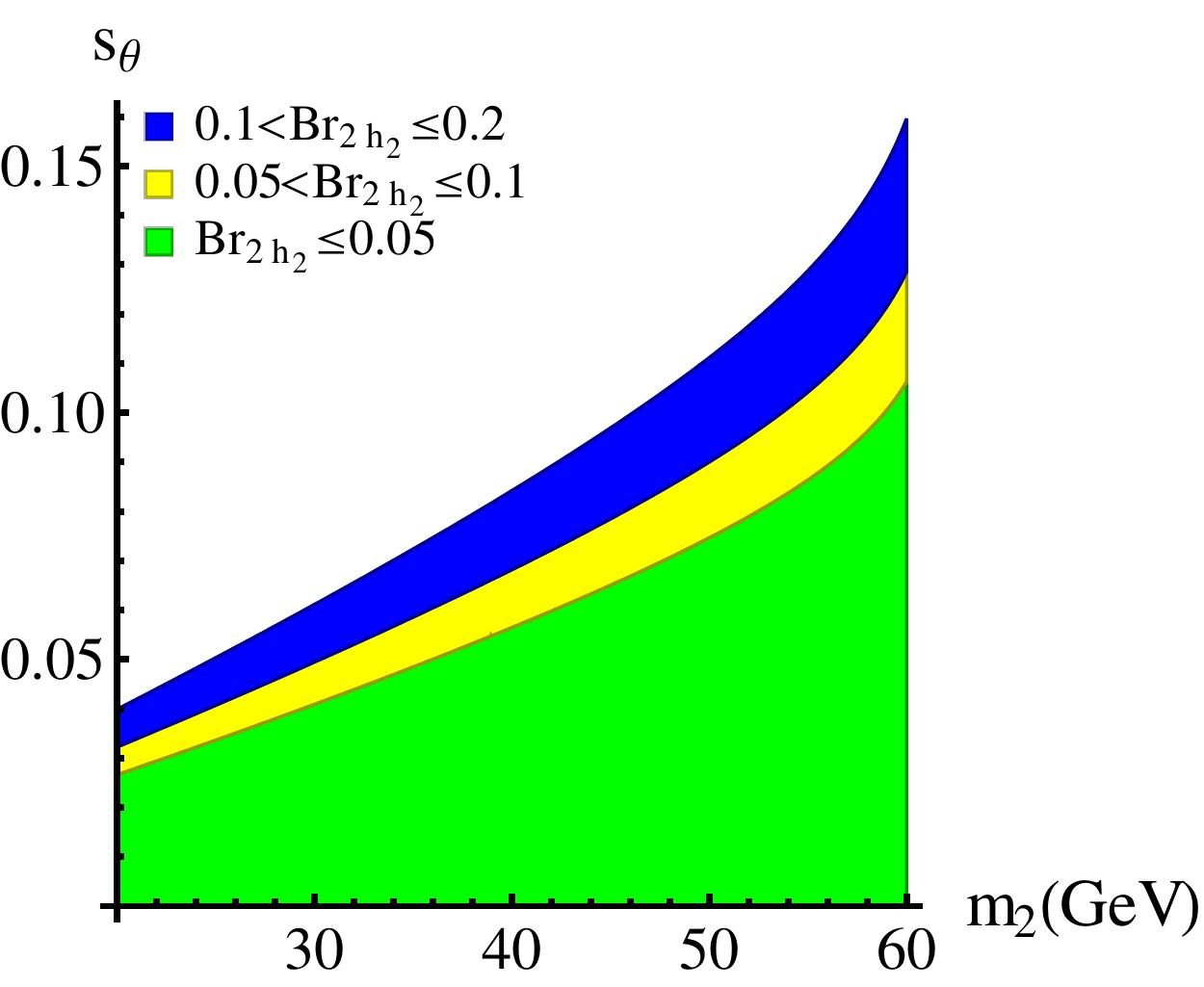}\includegraphics[scale=0.43]{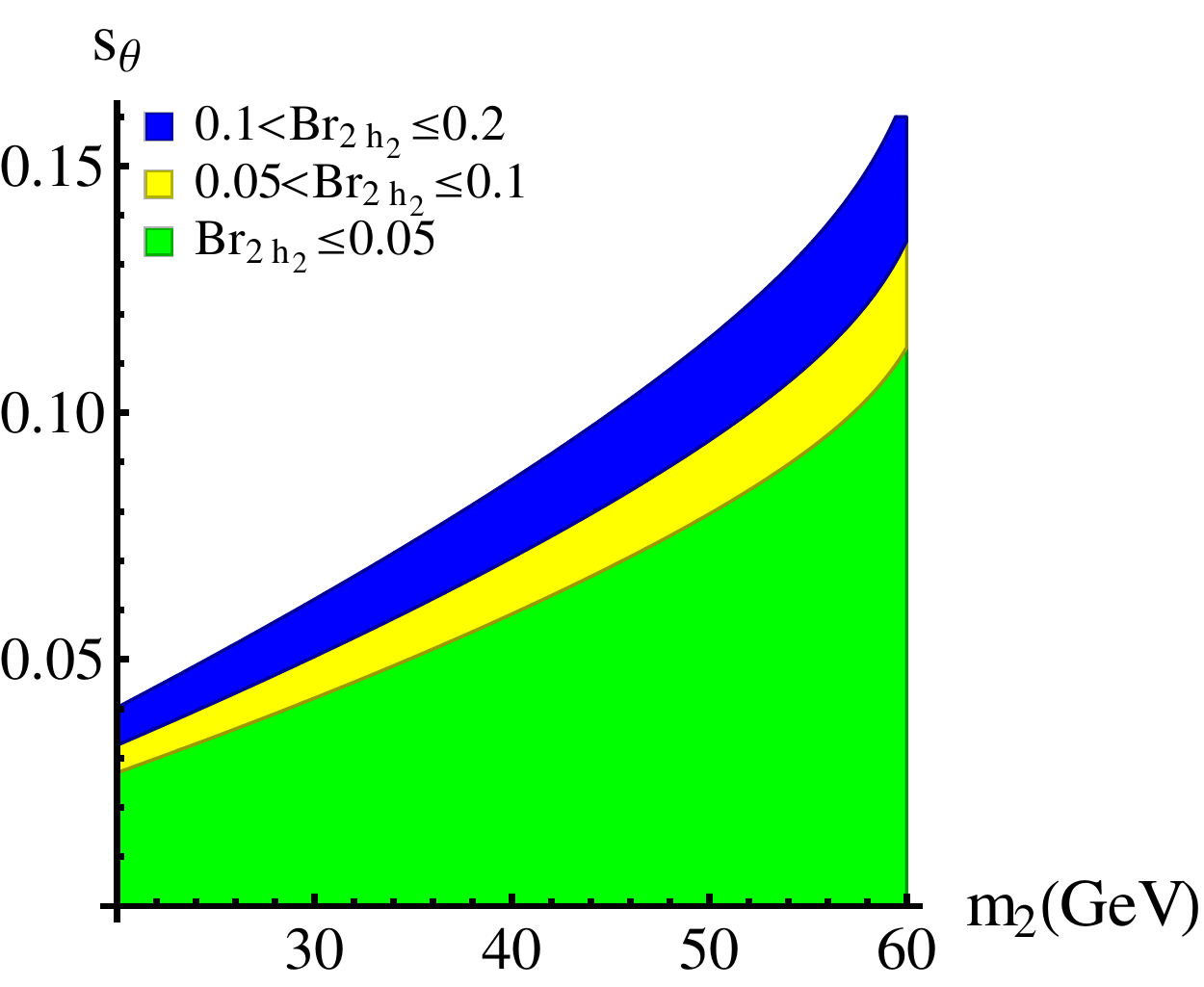}
\end{figure}

For $m_2\sim(15-80)~\textrm{GeV}$, experimentally, LEP direct searches through $e^+e^-\rightarrow Z^*\rightarrow Zh_2$
associated production process gave \cite{LEP}
\begin{equation}
s_{\theta}\lesssim(0.1-0.2)
\end{equation}
at $95\%$ C.L. assuming $\textrm{Br}_{h_2\rightarrow b\bar{b}}=1$. $\tilde{c}_{Zh_1h_2}$ cannot be constrained at LEP
since it is suppressed by a factor $(v/f)^3$. When $m_2<m_1/2$, it must face the $h_1$ rare decay constraint as well.
In SLH model, the dominant exotic decay channel is $h_1\rightarrow2h_2$ with a branching ratio $\textrm{Br}_{h_1\rightarrow2h_2}
\equiv\Gamma_{h_1\rightarrow2h_2}/\Gamma_1$. The partial decay width
\begin{equation}
\Gamma_{h_1\rightarrow2h_2}=\frac{\lambda_{122}^2f^2}{32\pi m_1}\sqrt{1-\frac{4m_2^2}{m^2_1}};
\end{equation}
while the $h_1$ total decay width
\begin{equation}
\Gamma_1=\Gamma_{h_1\rightarrow2h_2}+c^2_{\theta}\Gamma_{1,\textrm{SM}}.
\end{equation}
Based on the Higgs signal strengths measurements using full 2016 data \cite{stra,strc}, we perform a global-fit and obtain an estimation
\begin{equation}
\label{h12h2}
\textrm{Br}_{\textrm{exo}}\lesssim0.2\quad\textrm{and}\quad s_{\theta}\lesssim0.4,
\end{equation}
at $95\%$ C.L., which is a bit stricter than the previous constraint from LHC Run I \cite{exo}. We show the branching ratio distribution
in \autoref{EXO}. According to the figures, when $m_2\sim(20-60)~\textrm{GeV}$, we have $s_{\theta}\lesssim(0.04-0.16)$ which is a stricter
constraint than that from LEP direct searches. The numerical results are not sensitive to $f$ and $\beta$.

Theoretically, the allowed parameter region also depends on the details of EWSB, especially the contributions from cut-off scale,
$\delta m^2=-cf^2$. In the CP-violation case, (\ref{lammu}) becomes
\begin{equation}
\delta V'_h=2\epsilon f^2\left(2c_{\alpha}c_{2\alpha}c^2_{\xi}-c_{4\alpha}c_{2\xi}\right)\left(h^{\dag}h\right)+
\frac{\epsilon\left(c^2_{\alpha}c^2_{\xi}-2c_{2\alpha}c_{2\xi}\right)}{3s^2_{\beta}c^2_{\beta}}\left(h^{\dag}h\right)^2,
\end{equation}
leaving the other contributions to $\delta V_h$ unchanged. For $f=8~\textrm{TeV}$,
in the light $h_2$ scenario, $c$ is favored in the region $(0.01-0.02)$ since larger $c_2$ is excluded by the Higgs data. However, if
$c\lesssim0.01$, EWSB requires larger $s_{\theta}$ which was excluded by the Higgs rare decay constraints, thus smaller $c_2$ would
lead to the exclusion of light $h_2$ scenario. Larger $f$ requires smaller $c$, for example, if $f=12~\textrm{TeV}$, the lower limit
of $c$ reaches about $4\times10^{-3}$.

For a heavy $h_2$ (with $m_2\gtrsim200~\textrm{GeV}$), experimentally it is
constrained by LHC direct searches. At LHC, the gluon fusion process acquires dominant contribution through top quark
loop, and the amplitudes through heavy quark loops are suppressed by $(v/f)^2$, so thus $\sigma_{h_2}/\sigma_{h_2,\textrm{SM}}\approx
s^2_{\theta}$. If $m_2<2m_1$, the branching ratios of $h_2$ are the same as those of a SM-like Higgs boson with the mass $m_2$.
For $m_2>2m_1$, another decay channel $h_2\rightarrow 2h_1$ opens with a partial width
\begin{equation}
\Gamma_{h_2\rightarrow2h_1}=\frac{\lambda_{211}^2f^2}{32\pi m_2}\sqrt{1-\frac{4m_1^2}{m^2_2}}.
\end{equation}
Its branching ratio can reach $(20-30)\%$ when $m_2\gtrsim300~\textrm{GeV}$. If $m_2\gtrsim350~\textrm{GeV}$,
$h_2\rightarrow t\bar{t}$ decay channel can also open. Recently, ATLAS collaboration performed the direct searches through the channels
$pp\rightarrow h_2\rightarrow W^+W-,ZZ$ for $m_2>200~\textrm{GeV}$ with $36.1~\textrm{fb}^{-1}$ luminosity at $\sqrt{s}=13~\textrm{TeV}$
\cite{h2ZZ,h2WW}. If $m_2\lesssim1~\textrm{TeV}$, the strictest constraints come from the $h_2\rightarrow ZZ$ decay channel. Comparing
with the SM theoretical predictions \cite{hig12,LHCW}, we have a rough estimation
\begin{equation}
s_{\theta}\lesssim\left\{\begin{array}{ll}(0.1-0.4),&\textrm{for }m_2\sim(0.2-0.3)~\textrm{TeV};\\
0.2,&\textrm{for }m_2\sim(0.3-0.7)~\textrm{TeV};\\(0.2-0.4),&\textrm{for }m_2\sim(0.7-1)~\textrm{TeV}.\end{array}\right.
\end{equation}
at $95\%$ C.L. These constraints are a bit weaker than those in the light $m_2$ region.

Theoretical constraints here are similar to those in the case with light $h_2$. $c\sim(0.005-0.03)$ is favored in the heavy $h_2$ scenario.
In this scenario, the results are not sensitive to $f$ or $\beta$. Bound on $m_2$ is sensitive to $c$, but not sensitive to $s_{\theta}$,
which is different from the properties in light $h_2$ scenario.

\subsection{EDM Constraints}
The EDM effective interaction can be written as
\begin{equation}
\mathcal{L}_{\textrm{EDM}}=-\frac{\textrm{i}d_f}{2}\bar{f}\sigma^{\mu\nu}\gamma^5fF_{\mu\nu},
\end{equation}
which violated P- and CP-symmetries. In the SM, CP-violation comes only from complex CKM matrix so that the leading
contributions to the EDMs of electron and neutron arise at four- and three-loop level respectively. It is estimated
that \cite{EDM}
\begin{equation}
d_{e,\textrm{SM}}\sim10^{-38}~e\cdot\textrm{cm},\quad\quad d_{n,\textrm{SM}}\sim10^{-32}~e\cdot\textrm{cm},
\end{equation}
both of which are far below the recent experimental constraints \cite{EDMe,EDMn}
\begin{equation}
|d_e|<8.7\times10^{-29}~e\cdot\textrm{cm},\quad\quad|d_n|<3.0\times10^{-26}~e\cdot\textrm{cm},
\end{equation}
at $90\%$ C.L. However, in some BSM models, electron or neutron EDM can be generated at one- or two- loop level,
which means it may face strict experimental constraints.

\begin{figure}[h]
\caption{Dominant Feynman diagrams contributing to EDM of electron and quarks, and CEDM of quarks. $F$ running in the
loop includes $t,T,D,S$.}\label{EDMdiag}
\includegraphics[scale=0.9]{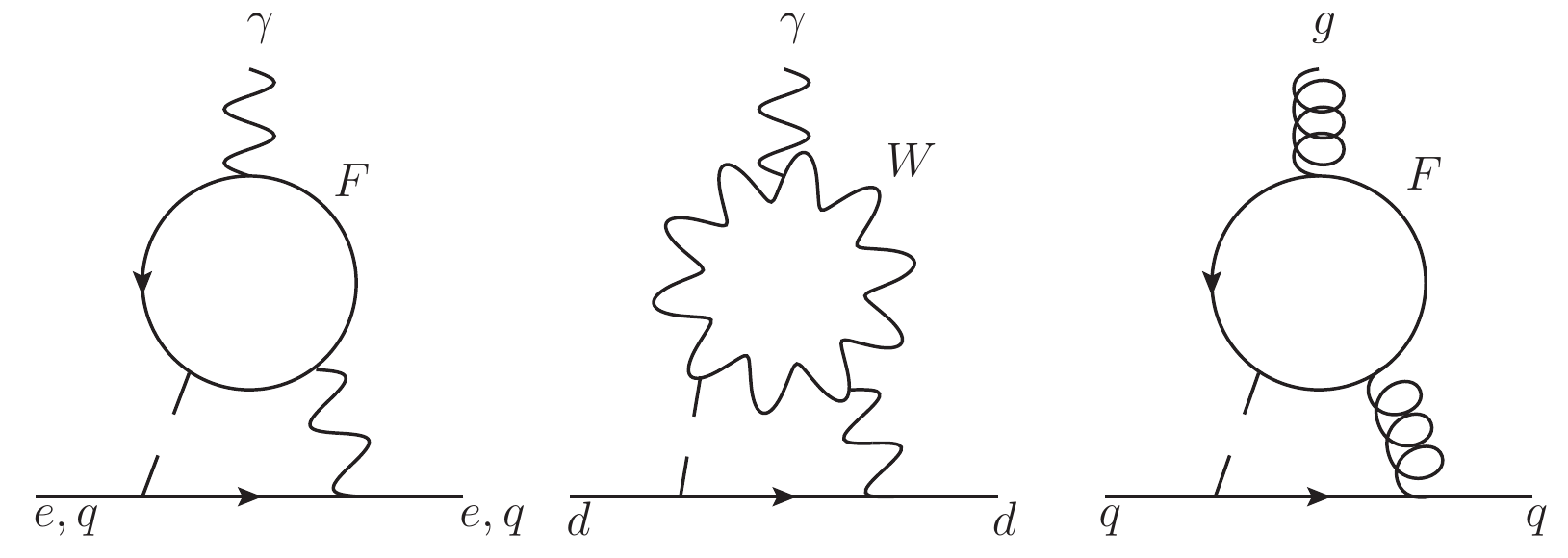}
\end{figure}

In the SLH model with spontaneous CP-violation, the leading contribution to electron EDM comes from the two-loop
``Barr-Zee" type diagrams \cite{BZ} with $F=t,T,D,S$ running in the loop, see the left diagram in \autoref{EDMdiag}.
Following the calculations in \cite{BZ,BZ2}, we have the analytical expression for the EDM of an electron as
\begin{equation}
\frac{d_e}{e}=\frac{3G_F\alpha_{\textrm{em}}m_es_{\theta}c_{\theta}}{(2\pi)^3}\left(\frac{v}{f}\right)\mathop{\sum}_{F}Q_F^2\delta_F
\frac{c_{2\beta}+c_{2\theta_{R,F}}}{s_{2\beta}}\left(g\left(\frac{m^2_F}{m_1^2}\right)-g\left(\frac{m^2_F}{m_2^2}\right)\right),
\end{equation}
in which the function
\begin{equation}
g(z)\equiv\frac{z}{2}\int_0^1dx\frac{1}{x(1-x)-z}\ln\frac{x(1-x)}{z}.
\end{equation}
Numerical results showed that $d_e$ is not sensitive to the masses of extra heavy quarks. For $0.2\lesssim t_{\beta}\lesssim8$, in the
whole mass region $m_2\sim(20-600)~\textrm{GeV}$, we have
\begin{equation}
\left|d_e\right|\lesssim8\times10^{-29}\left(\frac{8~\textrm{TeV}}{f}\cdot\frac{s_{2\theta}}{0.2}\right)~e\cdot\textrm{cm}\propto f^{-1}.
\end{equation}
The constraints from electron EDM are not strict due to the suppressions by $\theta$ and $f$.

For a neutron, its EDM comes from not only quarks' EDM, but also their color EDM (CEDM) operator \cite{EDM,BZ,BZ2}
\begin{equation}
\mathcal{O}_{\textrm{CEDM}}=-\frac{\textrm{i}g_s}{2}\tilde{d}_q\bar{q}_i\sigma^{\mu\nu}\gamma^5\left(t^a\right)_{ij}q_jG^a_{\mu\nu},
\end{equation}
where $\tilde{d}_q$ is the CEDM of the quark, $t^a$ denotes the color $\textrm{SU}(3)$ generator, and $i,j$ are color indices.
The $u$ quark EDM comes only from the left diagram in \autoref{EDMdiag}, just like that for electron; while the $d$ quark EDM
acquire contributions from both the left and middle diagrams in \autoref{EDMdiag}, because of the left-handed mixing between $d$
and $D$ quarks. The CEDM of quarks come from the right diagram in \autoref{EDMdiag}. Calculate at the EW scale, the quarks' EDM
and CEDM in the SLH model with spontaneous CP-violation are \cite{BZ2}
\begin{eqnarray}
d_u&=&-\frac{2m_u}{3m_e}d_e;\\
d_d&=&\frac{m_d}{3m_e}d_e+\frac{4G_F\alpha_{\textrm{em}}m_ds_{\theta}c_{\theta}}{9(2\pi)^3t_{\beta}}\left(\frac{v}{f}\right)
\left(f\left(\frac{m^2_t}{m^2_1}\right)-f\left(\frac{m^2_t}{m^2_2}\right)\right)\nonumber\\
&&-\frac{G_F\alpha_{\textrm{em}}m_ds_{\theta}c_{\theta}}{12(2\pi)^3t_{\beta}}\left(\frac{v}{f}\right)
\left[\left(\left(6+\frac{m_1^2}{m^2_W}\right)f\left(\frac{m^2_W}{m^2_1}\right)-
\left(6+\frac{m_2^2}{m^2_W}\right)f\left(\frac{m^2_W}{m^2_2}\right)\right)\right.\nonumber\\
&&\left.+\left(\left(10-\frac{m_1^2}{m^2_W}\right)g\left(\frac{m^2_W}{m^2_1}\right)-
\left(10-\frac{m_2^2}{m^2_W}\right)g\left(\frac{m^2_W}{m^2_2}\right)\right)\right];\\
\tilde{d}_u&=&-\frac{G_F\alpha_sm_us_{\theta}c_{\theta}}{2(2\pi)^3}\left(\frac{v}{f}\right)\mathop{\sum}_F\delta_F
\frac{c_{2\beta}+c_{2\theta_{R,F}}}{s_{2\beta}}\left(g\left(\frac{m^2_F}{m^2_1}\right)-g\left(\frac{m^2_F}{m^2_2}\right)\right);\\
\tilde{d}_d&=&\frac{m_d}{m_u}\tilde{d}_u-\frac{G_F\alpha_sm_ds_{\theta}c_{\theta}}{2(2\pi)^3t_{\beta}}\left(\frac{v}{f}\right)
\left(f\left(\frac{m^2_F}{m^2_1}\right)-f\left(\frac{m^2_F}{m^2_2}\right)\right);
\end{eqnarray}
in which the function
\begin{equation}
f(z)\equiv\frac{z}{2}\int_0^1dx\frac{1-2x(1-x)}{x(1-x)-z}\ln\frac{x(1-x)}{z}.
\end{equation}
After the running to hadron scale, the neutron EDM \cite{BZ2}
\begin{equation}
\frac{d_{n,\textrm{BZ}}}{e}\simeq0.63\frac{d_d}{e}+0.73\tilde{d}_d-0.16\frac{d_u}{e}+0.19\tilde{d}_u.
\end{equation}
Numerically, for $0.2\lesssim t_{\beta}\lesssim8$, in the whole mass region $m_2\sim(20-600)~\textrm{GeV}$, we have
\begin{equation}
\left|d_n\right|\lesssim1.4\times10^{-26}\left(\frac{8~\textrm{TeV}}{f}\cdot\frac{s_{2\theta}}{0.2}\right)~e\cdot\textrm{cm},
\end{equation}
which is still below the experimental limit. The constraint from neutron EDM is weaker than that from electron EDM.

\begin{figure}[h]
\caption{Additional Feynman diagrams contributing to neutron EDM.}\label{EDMadd}
\includegraphics[scale=0.9]{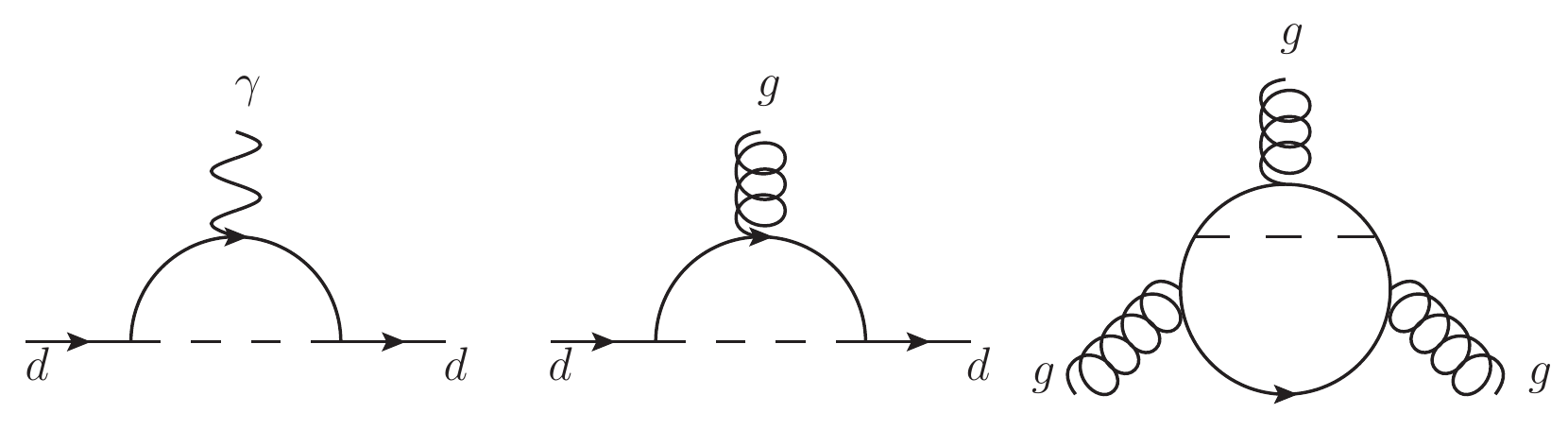}
\end{figure}

Besides the ``Barr-Zee" type diagram, there are also one-loop diagrams and Weinberg operator \cite{wein} contributing to neutron EDM,
see the Feynman diagrams in \autoref{EDMadd}.
Following (\ref{c1})-(\ref{c2}), we can estimate the one-loop contribution to neutron EDM (the left and middle diagrams in \autoref{EDMadd}) as
\begin{equation}
\left|\delta d_{n,\textrm{1-loop}}\right|\sim\frac{0.7s_{\theta}c_{\theta}}{32\pi^2t_{\beta}}
\frac{m_d\left|m^2_1-m^2_2\right|}{vfm^2_D}\ln\left(\frac{m^2_D}{\mu^2}\right),
\end{equation}
where the scale $\mu\sim\mathcal{O}(v)$. This result is sensitive to $m_D$ and $m_2$. For $m_D\sim\mathcal{O}(f)$ and $m_2\lesssim\mathcal{O}(v)$, $ \left|\delta d_{n,\textrm{1-loop}}\right|\lesssim\mathcal{O}(10^{-29}-10^{-27})~e\cdot\textrm{cm}$. The Weinberg operator (the right diagram in \autoref{EDMadd}) \cite{wein},
\begin{equation}
\mathcal{O}_W=-\frac{w}{3}f^{abc}G_{\mu\nu}^aG^{\nu,b}_{\rho}\tilde{G}^{\mu\rho,c}
\end{equation}
in which $f^{abc}$ is the structure constant of $\textrm{SU}(3)$ group, contribute to neutron EDM as \cite{BZ2}
\begin{equation}
\frac{\delta d_{n,W}}{e}\simeq(9.8~\textrm{MeV})w.
\end{equation}
In the SLH model with spontaneous CP-violation, we have
\begin{equation}
w=\frac{G_F\alpha_s}{(4\pi)^3}\left(\frac{v}{f}\right)s_{\theta}c_{\theta}\frac{c_{2\theta_{R,t}}+c_{2\beta}}{s_{2\beta}}
\left(W\left(\frac{m^2_t}{m^2_2}\right)-W\left(\frac{m^2_t}{m^2_1}\right)\right),
\end{equation}
where the function \cite{BZ2}
\begin{equation}
W(z)\equiv z^2\int^1_0du\int^1_0dv\frac{(1-v)(uv)^3}{((1-u)(1-v)+v(1-uv)z)^2}.
\end{equation}
Typical $|\delta d_{n,W}|\lesssim\mathcal{O}(10^{-28})~e\cdot\textrm{cm}$. Thus we can conclude that for neutron EDM, the
contributions from \autoref{EDMadd} are sub-dominant.

There are also upper limits on heavy atoms' EDM. The recent measurement on $^{199}$Hg atom's EDM set new limit
$|d_{\textrm{Hg}}|<7.4\times10^{-30}~e\cdot\textrm{cm}$ at $95\%$ C.L. \cite{dHg} which provides an indirect constraint
$d_n\lesssim1.6\times10^{-26}~e\cdot\textrm{cm}$ \cite{dnind}. The SLH model with spontaneous CP-violation is still allowed by this
new indirect constraints. The theoretical estimation on the EDM of Hg contains rather large uncertainties \cite{atomEDM}, thus
it cannot directly provide further constraint on this model.

\section{Future Collider Tests of the CP-violation Effects}
\label{test}
Recent Higgs data have already confirmed the $0^+$ component of $h_1$ \cite{higCP}. Following the idea in \cite{mao,testCPV},
we should try to measure tree level $h_2VV$ and $h_1h_2V$ vertices to confirm CP-violation in the scalar sector. For different
$h_2$ mass, we need different future colliders.
\subsection{Measuring $h_2VV$ Vertex}
If $h_2$ is light (for example, $m_2\ll v$), it is difficult to be discovered at LHC because of the large QCD backgrounds at
low mass region. To test this scenario, we need future $e^+e^-$ colliders. For example, at CEPC \cite{CEPC} or TLEP \cite{TLEP}
with $\sqrt{s}\sim(240-250)~\textrm{GeV}$, the $h_2VV$ vertex can be measured through the $e^+e^-\rightarrow Z^*\rightarrow Zh_2$
associated production process. Its cross section is \cite{LEP,sig}
\begin{equation}
\sigma_{Zh_2}=\frac{\pi\alpha^2_{\textrm{em}}\left(8s^4_{W}-4s^2_{W}+1\right)\cdot s^2_{\theta}}{96s(1-m^2_Z/s)^2s^4_{W}c^4_{W}}\left(\mathcal{F}^3\left(\frac{m_Z^2}{s},\frac{m^2_2}{s}\right)+
\frac{12m^2_Z}{s}\mathcal{F}\left(\frac{m_Z^2}{s},\frac{m^2_2}{s}\right)\right),
\end{equation}
in which the function
\begin{equation}
\mathcal{F}(x,y)\equiv\sqrt{1+x^2+y^2-2x-2y-2xy}.
\end{equation}
With $5~\textrm{ab}^{-1}$ luminosity at CEPC, the inclusive discovery potential on $s_{\theta}$ can reach $5\sigma$ if
$s_{\theta}\sim0.15$ at low mass region ($m_2\lesssim70~\textrm{GeV}$) \cite{testCPV} through the ``recoil mass" technique \cite{CEPC,rec,rec2}.
This result does not depend on the decay channel of $h_2$, and it is not sensitive to $m_2$ in this region. With the the help of ``$p_T$ balance
cut" method \cite{ptb} to reduce large backgrounds with photons, the $5\sigma$ discovery bound on $s_{\theta}$ can reach about $0.1$ with a tiny
breaking of inclusiveness. If we completely give up the inclusiveness in this measurement and consider only the $h_2\rightarrow b\bar{b}$ decay
channel, the $5\sigma$ discovery bound on $s_{\theta}$ can be suppressed to about $(4-5)\times10^{-2}$ according to \cite{testCPV} \footnote{Simulation
details about the cross sections of the background channels were not shown in the text of \cite{testCPV}.}. This result means the allowed regions
obtained in \autoref{EXO} are still possible to be discovered at $5\sigma$ level at CEPC with $5~\textrm{ab}^{-1}$ luminosity. For larger $m_2$
when it is close to $Z$-peak, large $ZZ$ background will decrease the sensitivity on $s_{\theta}$ measured though this channel.

For light $h_2$, we can also measure $s_{\theta}$ through $Z\rightarrow Z^*(f\bar{f})h_2$ rare decay, if an $e^+e^-$ collider runs at
$Z$-pole ($\sqrt{s}=m_Z$). The branching ratio \cite{PDG}
\begin{equation}
\textrm{Br}_{Z\rightarrow Z^*h_2}=\frac{s^2_{\theta}}{\pi^2m_Z}\int_0^{\pi}\sin\phi d\phi\int_0^{m_Z-m_2}dq
\frac{q^3p_2}{\left(q^2-m^2_Z\right)^2}\left(2+\frac{m^2_2\beta^2\sin^2\phi}{1-\beta^2}\right),
\end{equation}
where $q$ is the invariant mass of $Z^*$. The momentum of $h_2$ in initial $Z$ frame and the relative velocity between
$h_2$ and $Z^*$ are respectively
\begin{eqnarray}
p_2&=&\frac{\sqrt{\left(m^2_Z-\left(m_2-q\right)^2\right)\left(m^2_Z-\left(m_2+q\right)^2\right)}}{2m_Z},\\
\beta&=&\frac{m_Zp_2}{p_2^2+\sqrt{(p_2^2+m_2^2)(p^2_2+q^2)}}.
\end{eqnarray}
With $10^{12}$ $Z$-boson events as the goal of a ``Tera-$Z$" factory, the typical sensitivity to this rare decay branching ratio
is about $(10^{-8}-10^{-7})$ \cite{Zrare}, which means it has a better sensitivity to discover nonzero $s_{\theta}$ comparing
with the $Zh_2$ associated production channel in the whole mass region $m_2\lesssim70~\textrm{GeV}$.

For a heavy $h_2$ (for example, $m_2\sim\mathcal{O}(v)$), LHC future direct searches will discover it or set a stricter limit on
$s_{\theta}$, through its $ZZ$ decay channel \cite{CMSFUT}. Through merely visible leptonic decay channel, with $3~\textrm{ab}^{-1}$,
the $5\sigma$ discovery bounds would be around $s_{\theta}\sim(0.1-0.2)$, which is similar to the current upper limits using the
combination of $h_2\rightarrow4\ell$ and $h_2 \rightarrow2\ell2\nu$ channels \cite{LHCW,h2ZZ,CMSFUT}. We also expect the $2\ell2\nu$
channel can help to increase the sensitivity on
$s_{\theta}$ at future LHC. When $m_2\gtrsim0.6~\textrm{TeV}$, the $2\ell2\nu$ channel would become more sensitive than the $4\ell$
channel \cite{h2ZZ}.
\subsection{Measuring $h_1h_2V$ Vertex}
Based on the improved formalism of SLH model \cite{Zh1h2}, we obtain the $Zh_1h_2$ vertex in (\ref{Vh1h2}). $\tilde{c}^{as}_{Zh_1h_2}$ is
suppressed by a factor $(v/f)^3\lesssim\mathcal{O}(10^{-5})$ and thus the associated production channels cannot be used to measure this vertex.
Similarly, precision measurements on $h_1\rightarrow Z^{(*)}h_2$ are also useless to test this vertex, since the typical $5\sigma$
discovery bounds for such rare decay channels are of $\mathcal{O}(10^{-3})$ \cite{CEPC,rare}. That means we must turn to the heavy
neutral gauge boson sector for help.

According to (\ref{Vh1h2}), $\tilde{c}^{as}_{Z'h_1h_2}$ is suppressed by a factor $(v/f)$, and there is no suppression in
$\tilde{c}^{as}_{Yh_1h_2}$. These vertices will become helpful to confirm the $0^-$ component in at least one of the scalars.
Since $m_{Z'}\gg m_{1,2}$, the decay branching ratio
\begin{equation}
\textrm{Br}_{Z'\rightarrow h_1h_2}=\frac{m^3_{Z'}}{48\pi\Gamma_{Z'}f^2}\left(\frac{v}{ft_{2\beta}}\right)^2.
\end{equation}
Assuming the heavy quark masses $m_F>m_{Z'}/2$ thus $\tilde{Z}'\rightarrow F\bar{F}$ decay channels cannot be opened. The
total width $\Gamma_{Z'}\approx6.5\times10^{-3}f$ if we choose the ``anomaly free" embedding \cite{SLHi}. Numerically, we have
\begin{equation}
\textrm{Br}_{Z'\rightarrow h_1h_2}\simeq1.7\times
10^{-4}\left(\frac{8~\textrm{TeV}}{ft_{2\beta}}\right)^2\propto f^{-2}.
\end{equation}
When $\beta\sim\pi/4$, this decay channel vanishes, while if $\beta$ is close to $0$ or $\pi/2$, there is an enhancement by $t^{-2}_{2\beta}$.
It decreases quickly when $f$ increases.

For this process, we need future $pp$ colliders with larger $\sqrt{s}$, for example, $(50-100)~\textrm{TeV}$ \cite{CEPC,100T}. Since at
LHC, when $m_{Z'}\gtrsim5~\textrm{TeV}$, the event number of $pp\rightarrow\tilde{Z}'\rightarrow h_1h_2$ cannot reach $\mathcal{O}(1)$ with
$3~\textrm{ab}^{-1}$ luminosity \cite{100T}. However, with the same luminosity at $\sqrt{s}=100~\textrm{TeV}$ $pp$ collider, the events number
can reach $N_{pp\rightarrow Z'\rightarrow h_1h_2}\sim\mathcal{O}(10^2-10^3)$ for $m_{Z'}\sim5~\textrm{TeV}$, and $N_{pp\rightarrow Z'
\rightarrow h_1h_2}\sim\mathcal{O}(10-10^2)$ for $m_{Z'}\sim10~\textrm{TeV}$ \cite{100T}. This implies the $Z'h_1h_2$ vertex in the SLH model is
testable at $\sqrt{s}=100~\textrm{TeV}$ $pp$ collider.

If we can discover nonzero values for both $h_2ZZ$ and $Z'h_1h_2$ couplings, we can confirm the CP-violation effects in the scalar sector.

\section{Conclusions and Discussions}
\label{conc}
We proposed the possibility of spontaneous CP-violation in the scalar sector of the SLH model in this paper.
Through adding a new interaction term, $\epsilon\left(\Phi_1^{\dag}\Phi_2\right)^2+\textrm{H.c.}$, in the scalar
potential, the pseudoscalar field $\eta$ can acquire a nonzero VEV which means CP-violation happens spontaneously.
Both scalars then become CP-mixing states. In this paper, we denote $h_1$ as the SM-like Higgs boson with its mass
$m_1=125~\textrm{GeV}$, and $h_2$ is the extra scalar. Based on the improved
SLH formalism (see \autoref{form} in the appendix), we derived the interactions in this model.

Facing strict experimental constraints, the spontaneous CP-violation SLH model is still not excluded. LHC Run II data
have already push the lower limit of the scale $f$ to about $7.5~\textrm{TeV}$, which means the EW precision tests
only provide sub-dominant constraints on $f$. For the extra scalar $h_2$, we have two scenarios based on its mass,
$m_2\sim\mathcal{O}(v)$ or $m_2\ll v$. For a light $h_2$, the most strict constraint comes from $h_1\rightarrow2h_2$
rare decay channel. The $95\%$ C.L. upper limit on $s_{\theta}$ is $(0.04-0.16)$ for $m_2\sim(20-60)~\textrm{GeV}$.
While for large $m_2\sim\mathcal{O}(v)$, the $95\%$ C.L. upper limit on $s_{\theta}$ varies in the region $(0.1-0.4)$,
especially when $m_2\sim(300-700)~\textrm{GeV}$, the $95\%$ C.L. upper limit on $s_{\theta}$ is about $0.2$. In both
scenarios, tiny but nonzero contributions from cut-off scale are necessary. As a CP-violation model, it must also face
the EDM constraints. Since the effects are suppressed by $s_{\theta}v/f$, the constraints are weak. The most strict EDM
constraint comes from electron, which favors $0.2\lesssim t_{\beta}\lesssim8$ in the whole $m_2\sim(20-600)~\textrm{GeV}$
mass region.

We also discussed the future collider tests of this model. The basic idea is to discover nonzero $h_2VV$ and $Vh_1h_2$
vertices. For a light $h_2$ , we can test $h_2ZZ$ vertex at future $e^+e^-$ colliders, as Higgs factories or $Z$-factory.
With $5~\textrm{ab}^{-1}$ at CEPC, for $m_2\lesssim70~\textrm{GeV}$, $s_{\theta}\sim(4-5)\times10^{-2}$ can be discovered
at $5\sigma$ level; while with $10^{12}$ $Z$-boson events at $Z$-pole, we can have a better sensitivity in the same mass
region. For a heavy $h_2$ with $m_2\sim\mathcal{O}(v)$, the vertex can be tested through $gg\rightarrow h_2\rightarrow ZZ$
channel at LHC. With $3~\textrm{ab}^{-1}$ luminosity, the $5\sigma$ discovery bound is around $(0.1-0.2)$ through merely the
$4\ell$ decay channel. The $2\ell2\nu$ decay channel is also expected to help increase the sensitivity on $s_{\theta}$,
especially in large $m_2$ region. Based on the improved formalism, we know the $Zh_1h_2$ vertex is suppressed by $(v/f)^3$,
thus we must ask a heavy gauge boson, such as $Z'$, for help. Since $\textrm{Br}_{Z'\rightarrow h_1h_2}\lesssim\mathcal{O}(10^{-4}-10^{-3})$,
it is difficult to be tested at LHC. We need $pp$ colliders with larger $\sqrt{s}$. For example, if $\sqrt{s}=100~\textrm{TeV}$,
with $3~\textrm{ab}^{-1}$ luminosity, we can obtain $\mathcal{O}(10^2-10^4)$ events for $pp\rightarrow \tilde{Z}'\rightarrow h_1h_2$ process
in the mass region $m_{Z'}\sim(5-10)~\textrm{TeV}$ which means it may become testable. CP-violation in the scalar sector will be
confirmed if both nonzero $h_2ZZ$ and $Z'h_1h_2$ vertices are discovered.

This model is attractive both theoretically and phenomenologically. Theoretically, in this model, we proposes a new possible
CP-violation source, which may provide new understanding of the matter-antimatter asymmetry problem in the Universe. Besides this,
the spontaneous CP-violation mechanism is also a possible solution to the strong-CP problem, which is worthy to study further. This
model is also a candidate to connect between the composite Higgs mechanism and CP-violation in the scalar sector. Based on this, new
CP-violation effects are naturally suppressed by the global symmetry breaking scale $f$, as shown in the calculation of electron and
neutron EDM.

Phenomenologically, it is an application of the basic idea to measure $h_2VV$ and $Vh_1h_2$ vertices. It provides an example to
show how extra scalars and gauge bosons can help to confirm new CP-violation sources, which also implies the importance to search
for $VVS$- and $VSS$-type vertices. It also shows another motivation for future $e^+e^-$ and $pp$ colliders.

\section*{Acknowledgement}
We thank Jordy de Vries, Shi-ping He, Fa-peng Huang, Gang Li, Jia Liu, Lian-tao Wang, Ke-pan Xie, Ling-xiao Xu, Wen Yin, Felix Yu, Chen Zhang,
and Shou-hua Zhu for helpful discussions. This work was partly supported by the China Postdoctoral Science Foundation (Grant No. 2017M610992).

\numberwithin{equation}{section}
\appendix
\section{Improved Formalism of the SLH Model}
\label{form}
In this section show the improved formalism for the SLH model based on \cite{Zh1h2}. The neutral scalar sector (including six degrees of freedom)
can be divided into CP-even and CP-odd parts. The CP-odd part, denoting as $G_i$ running over $\eta$, $G$, $G'$, and $y^2$, is not canonically-normalized.
We can write the kinetic term as
\begin{equation}
\mathcal{L}\supset\frac{1}{2}\mathbb{K}_{ij}\partial^{\mu}G_i\partial_{\mu}G_j.
\end{equation}
The matrix elements of $\mathbb{K}$ are calculated to $\mathcal{O}\left((v/f)^3\right)$ in \cite{Zh1h2}.
If we rewrite this term in another basis $S_i=U_{ij}G_j$ which is canonically-normalized
\begin{equation}
\mathcal{L}\supset\frac{1}{2}\delta_{ij}\partial^{\mu}S_i\partial_{\mu}S_j
\end{equation}
thus we can define a inner product $\langle S_i|S_j\rangle=\delta_{ij}$ in the linear space spanned by the scalars $S_i$. A straightforward calculation
shows that
\begin{equation}
\langle G_i|G_j\rangle=\left(\mathbb{K}^{-1}\right)_{ij}.
\end{equation}

The VEVs in $\Phi_{1,2}$ will lead to two-point transitions between gauge bosons and pseudo-scalars as
\begin{equation}
\label{twop}
\mathcal{L}\supset V_p^{\mu}\mathbb{F}_{pi}\partial_{\mu}G_i
\end{equation}
where $V_p$ denotes a gauge boson running over $Z$, $Z'$, and $Y^2$, and $\mathbb{F}$ is a $4\times3$ matrix. The
matrix elements of $\mathbb{F}$ are also calculated to $\mathcal{O}\left((v/f)^3\right)$ in \cite{Zh1h2}.
The gauge fixing term must provide the two-point transition like
\begin{equation}
\label{GF}
\mathcal{L}_{\textrm{G.F.}}\supset\left(\partial_{\mu}V_p^{\mu}\right)\mathbb{F}_{pi}G_i
\end{equation}
to cancel all contributions from (\ref{twop}). Define
\begin{equation}
\bar{G}_p=\mathbb{F}_{pi}G_i,
\end{equation}
in the convention of \cite{SLH3} (which is also the convention of this paper), we can derive that
\begin{equation}
\langle\eta|\bar{G}_p\rangle=0,\quad\textrm{and}\quad\langle\bar{G}_p|\bar{G}_q\rangle=\left(\mathbb{M}^2_V\right)_{pq},
\end{equation}
through a straightforward calculation where $\mathbb{M}^2_V$ is the mass matrix for gauge bosons in the basis $(Z,Z',Y^2)$.
Calculate to the leading order of $(v/f)$ for every matrix element, we have
\begin{equation}
\mathbb{M}^2_V=g^2\left(\begin{array}{ccc}\frac{v^2}{4c^2_W}&\frac{c_{2W}v^2}{4c^3_W\sqrt{3-t^2_W}}&\frac{v^3}{3\sqrt{2}c_Wt_{2\beta}f}\\
\frac{c_{2W}v^2}{4c^3_W\sqrt{3-t^2_W}}&\frac{2f^2}{3-t^2_W}&\frac{v^3}{3\sqrt{6-t^2_W}c^2_Wt_{2\beta}f}\\
\frac{v^3}{3\sqrt{2}c_Wt_{2\beta}f}&\frac{v^3}{3\sqrt{6-t^2_W}c^2_Wt_{2\beta}f}&\frac{f^2}{2}\end{array}\right).
\end{equation}
Using an orthogonal matrix $\mathbb{R}$, we can diagonalize $\mathbb{M}^2_V$ as
\begin{equation}
\left(\mathbb{R}\mathbb{M}^2_V\mathbb{R}^T\right)_{pq}=m^2_p\delta_{pq},\quad\textrm{and}\quad\tilde{V}_p=\mathbb{R}_{pq}V_q,
\end{equation}
where $\tilde{V}_p$ denotes the mass eigenstate of a gauge boson and $m_p$ is its mass. The matrix elements of $\mathbb{R}$ are calculated to $\mathcal{O}\left((v/f)^3\right)$ in \cite{Zh1h2} as well. For simplify, to this order, the off-diagonal elements can also be expressed as
\begin{equation}
\mathbb{R}_{pq}=\frac{\left(\mathbb{M}^2_V\right)_{pq}}{\left(\mathbb{M}^2_V\right)_{pp}-\left(\mathbb{M}^2_V\right)_{qq}}.
\end{equation}
It is natural for us to define
\begin{equation}
\tilde{G}_p\equiv\frac{\mathbb{R}_{pq}\bar{G}_q}{m_p}=\frac{\left(\mathbb{RF}\right)_{pi}G_i}{m_p}.
\end{equation}
According to $\langle\eta|\eta\rangle=\left(\mathbb{K}^{-1}\right)_{11}\approx1+(2/t^2_{2\beta})(v/f)^2$, we should also define
\begin{equation}
\tilde{\eta}\equiv\frac{\eta}{\sqrt{\left(\mathbb{K}^{-1}\right)_{11}}}.
\end{equation}
It is easy to check that in the basis $\left(\tilde{\eta},\tilde{G}_p\right)$, the kinetic part is canonically-normalized.
(\ref{GF}) also becomes $m_p\left(\partial_{\mu}\tilde{V}_p^{\mu}\right)\tilde{G}_p$, thus it is natural to choose the gauge
fixing term as
\begin{equation}
\mathcal{L}_{\textrm{G.F.}}=-\mathop{\sum}_p\frac{1}{2\xi_p}\left(\partial_{\mu}\tilde{V}_p^{\mu}-\xi_pm_p\tilde{G}_p\right)^2.
\end{equation}
It is now clear that $\tilde{G}_p$ is the corresponding Goldstone eaten by $\tilde{V}_p$, and its mass should be $\sqrt{\xi_p}m_p$ where $\xi_p$ is the corresponding gauge parameter. To this step, we have already built the formalism to treat a model with non-canonically-normalized scalar sector and the SLH model is one of the examples. The main point is that all the two-point transitions must be carefully canceled if we don't want these kind of Feynman diagrams appear during calculation.

Because of the $\eta$ components in the Goldstone fields, the interactions including $\eta$ must be changed comparing with the
naively calculated case. We divide $\mathbb{F}$ into
\begin{equation}
\mathbb{F}\equiv\left(\tilde{f},\tilde{\mathbb{F}}\right)
\end{equation}
where $\tilde{f}_p=\mathbb{F}_{p1}$ is a $1\times3$ vector and $\tilde{\mathbb{F}}$ is a $3\times3$ matrix. Thus for any kind of
couplings including the pseudo-scalar degrees of freedom, if we write the coefficients as $\left(c_{\eta},c_j\right)$ in $G_i$ basis
where $c_j$ runs for the couplings including $G$, $G'$ and $y^2$, the physical coupling should be
\begin{equation}
\label{ceta}
\tilde{c}_{\eta}=\sqrt{\left(\mathbb{K}^{-1}\right)_{11}}\left(c_{\eta}-c_j\left(\tilde{\mathbb{F}}^{-1}\tilde{f}\right)_j\right).
\end{equation}
For example, the anti-symmetric type $Vh\eta$ couplings in mass eigenstates can be parameterized as
\begin{equation}
\mathcal{L}_{Vh\eta}=\frac{g}{2}\left(h\partial^{\mu}\eta-\eta\partial^{\mu}h\right)\left(\tilde{c}^{as}_{Zh\eta}\tilde{Z}_{\mu}
+\tilde{c}^{as}_{Z'h\eta}\tilde{Z}'_{\mu}+\tilde{c}^{as}_{Yh\eta}\tilde{Y}^2_{\mu}\right).
\end{equation}
With the improved formalism, we can calculate to the leading order of $(v/f)$ as
\begin{equation}
\tilde{c}^{as}_{Zh\eta}=\frac{1}{2\sqrt{2}c^3_Wt_{2\beta}}\left(\frac{v}{f}\right)^3,\quad\tilde{c}^{as}_{Z'h\eta}=
\frac{2\sqrt{2}}{\sqrt{3-t^2_W}t_{2\beta}}\left(\frac{v}{f}\right),\quad\tilde{c}^{as}_{Yh\eta}=-1.
\end{equation}
The first two results are quite different from those appearing in previous papers \cite{SLH2,SLH3}. Similarly, the Yukawa couplings between
$\eta$ and SM fermions can be parameterized as
\begin{equation}
\mathcal{L}_{\eta f\bar{f}}=-\mathop{\sum}_{f}c_{\eta,f}\frac{\textrm{i}m_f}{v}\bar{f}\gamma^5f\eta.
\end{equation}
According to (\ref{ceta}), $c_{\eta,f}=0$ to all order of $(v/f)$ for $f=\nu,\ell,u,c,b$. This result is also quite
different from that in previous papers \cite{SLH2,SLH3}. For $f=t,d,s$, to the leading order of $(v/f)$, we have
\begin{equation}
c_{\eta,f}=-\delta_f\left(\frac{v}{\sqrt{2}f}\right)\frac{c_{2\beta}+c_{2\theta}}{s_{2\beta}}
\end{equation}
which is generated by the left-handed mixing between SM fermion and additional heavy fermion. Formally all these results
can be calculated to all order of $(v/f)$, though some of the results are extremely lengthy.

\clearpage\end{CJK*}


\begin{thebibliography}{99}

\bibitem{mass}The ATLAS and CMS Collaborations, Phys. Rev. Lett. 114, 191803, (2015).
\bibitem{PDG}K. A. Olive {\em et al.} (Particle Data Group), Chin. Phys. C 38, 090001 (2014); Chin. Phys. C 40, 100001 (2016).
\bibitem{higdisc}The ATLAS Collaboration, Phys. Lett. B 716, 1 (2012); The CMS Collaboration, Phys. Lett. B 716, 30 (2012).
\bibitem{stra}The ATLAS Collaboration, Report No. ATLAS-CONF-2017-045; Report No. ATLAS-CONF-2017-043; Report No. ATLAS-CONF-2017-041.
\bibitem{strc}The CMS Collaboration, Report No. CMS-PAS-HIG-16-044; Report No. CMS-PAS-HIG-16-041; Report No. CMS-PAS-HIG-16-040.
\bibitem{LH}N. Arkani-Hamed, A. G. Cohen, and H. Georgi, Phys. Lett. B 513, 232 (2001); N. Arkani-Hamed, A. G. Cohen, E. Katz, and A. E. Nelson,
 J. High Energy Phys. 07, 034 (2002); N. Arkani-Hamed, A. G. Cohen, E. Katz, A. E. Nelson, T. Gregoire, and J. G. Wacker,
 J. High Energy Phys. 08, 021 (2002); M. Schmaltz and D. Tucker-Smith, Ann. Rev. Nucl. Part. Sci. 55, 229 (2005).
\bibitem{comp}D. B. Kaplan and H. Georgi, Phys. Lett. B 136, 183 (1984).
\bibitem{SLH}D. E. Kaplan and M. Schmaltz, J. High Energy Phys. 10, 039 (2003); M. Schmaltz, J. High Energy Phys. 08, 056 (2004).
\bibitem{SLHi}T. Han, H. E. Logan, and L.-T. Wang, J. High Energy Phys. 01, 099 (2006).
\bibitem{SLH2}W. Kilian, D. Rainwater, and J. Reuter, Phys. Rev. D 71, 015008 (2005); Phys. Rev. D 74, 095003 (2006);
 Phys. Rev. D 74, 099905 (2006, erratum); K. Cheung and J. Song, Phys. Rev. D 76, 035007 (2007);
 K. Cheung, J. Song, P. Tseng, and Q.-S. Yan, Phys. Rev. D 78, 055015 (2008).
\bibitem{CPVdisc}J. H. Christenson, J. W. Cronin, V. L. Fitch, and R. Turlay, Phys. Rev. Lett. 13, 138 (1964).
\bibitem{KM}M. Kobayashi and T. Maskawa, Prog. Theor. Phys. 49, 652 (1973).
\bibitem{Cab}N. Cabibbo, Phys. Rev. Lett. 10, 531 (1963).
\bibitem{Plank}P. A. R. Ade {\em et al.} (The Planck Collaboration), Astron. Astrophys. 571, A16 (2014).
\bibitem{EWBG}D. E. Morrissey and M. J. Ramsey-Musolf, New J. Phys. 14, 125003 (2012).
\bibitem{EWBG2}A. G. Cohen, D. B. Kaplan, and A. E. Nelson, Phys. Lett. B 263, 86 (1991);
 Annu. Rev. Nucl. Part. Sci. 43, 27 (1993); J. Shu and Y. Zhang, Phys. Rev. Lett. 111, 091801 (2013).
\bibitem{2HDM}G. C. Branco, P. M. Ferreira, L. Lavoura, M. N. Rebelo, M. Sher, and J. P. Silva, Phys. Rep. 516, 1 (2012).
\bibitem{example1}L. Bento, G. C. Branco, and P. A. Parada, Phys. Lett. B 267, 95 (1991).
\bibitem{Lee}T. D. Lee, Phys. Rev. D 8, 1226 (1973); Phys. Rep. 9, 143 (1974).
\bibitem{example2}H. Georgi, Hadronic J. 1, 155 (1978).
\bibitem{example3}S. Weinberg, Phys. Rev. Lett. 37, 657, (1976).
\bibitem{higCP}The CMS Collaboration, Phys. Rev. D 89, 092007 (2014); Report No. CMS-PAS-HIG-17-011;
 The ATLAS Collaboration, Report No. ATLAS-CONF-2015-008.
\bibitem{strong}J. E. Kim and G. Garosi, Rev. Mod. Phys. 82, 557 (2010); S. M. Barr, Phys. Rev. Lett. 53, 329 (1984).
\bibitem{mao}Y.-N. Mao and S.-H. Zhu, Phys. Rev. D 90, 115024 (2014); Phys. Rev. D 94, 055008 (2016); Phys. Rev. D 94, 059904 (2016, erratum);
 Y.-N. Mao, PhD Thesis (Peking University, 2016).
\bibitem{NMC}B. Gripaios, A. Pomarol, F. Riva, and J. Serra, J. High Energy Phys. 04, 070 (2009).
\bibitem{CPVLH}Z. Surujon and P. Uttayarat, Phys. Rev. D 83, 076010 (2011); H. E. Haber and Z. Surujon, Phys. Rev. D 86, 075007 (2012).
\bibitem{strongmao}Y.-N. Mao, in preparation.
\bibitem{EDM}M. Pospelov and A. Ritz, Ann. Phys. (Amsterdam) 318, 119 (2005).
\bibitem{modify}A. Hocker and Z. Ligeti, Ann. Rev. Nucl. Part. Sci. 56, 501 (2006); J. Charles, S. Descotes-Genon,
 Z. Ligeti, S. Monteil, M. Papucci, and K. Trabelsi, Phys. Rev. D 89, 033016 (2014).
\bibitem{anoZ}B. Grzadkowski, O. M. Ogreid, and P. Osland, J. High Energy Phys. 11, 084 (2014); PoS CORFU2014, 086 (2015).
\bibitem{dist}S. Berge, W. Bernreuther, and J. Ziethe, Phys. Rev. Lett. 100, 171605 (2008);
 S. Berge, W. Bernreuther, and S. Kirchner, Phys. Rev. D 92, 096012 (2015).
 S. Berge, W. Bernreuther, and H. Spiesberger, Phys. Lett. B 727, 488 (2013).
 P. S. Bhupal Dev, A. Djouadi, R. M. Godbole, M. M. M$\ddot{\textrm{u}}$hlleitner, and S. D. Rindani, Phys. Rev. Lett. 100, 051801 (2008).
\bibitem{testCPV}G. Li, Y.-N. Mao, C. Zhang, and S.-H. Zhu, Phys. Rev. D 95, 035015 (2017).
\bibitem{K}A. M$\acute{\textrm{e}}$ndez and A. Pomarol, Phys. Lett. B 272, 313 (1991); J. F. Gunion and H. E. Haber, Phys. Rev. D 72, 095002 (2005).
\bibitem{Zh1h2}S.-P. He, Y.-N. Mao, C. Zhang and S.-H. Zhu, Phys. Rev. D 97, 075005 (2018).
\bibitem{SLH3}F. del $\acute{\textrm{A}}$guila, J. I. Illana, and M. D. Jenkins, J. High Energy Phys. 03, 080 (2011).
\bibitem{af}O. C. W. Kong, Report No. NCU-HEP-k009, arXiv: hep-ph/0307250; J. Korean Phys. Soc. 45, S404 (2004).
\bibitem{CW}S. R. Coleman and E. Weinberg, Phys. Rev. D 7, 1888 (1973).
\bibitem{SLHa}J. A. Casas, J. R. Espinosa, and I. Hidalgo, J. High Energy Phys. 03, 038 (2005).
\bibitem{obl1}M. E. Peskin and T. Takeuchi, Phys. Rev. Lett. 65, 964 (1990); Phys. Rev. D 46, 381 (1992).
\bibitem{obl2}M. Baak, J. Cuth, J. Haller, A. Hoecker, R. Kogler, K.M$\ddot{\textrm{o}}$nig, M. Schott, and J. Stelzer, Eur. Phys. J. C 74, 3046 (2014).
\bibitem{obl3}J. de Blas, M. Ciuchini, E. Franco, S. Mishima, M. Pierini, L. Reina, and L. Silvestrini, J. High Energy Phys. 12, 135 (2016).
\bibitem{SLH4}J. Reuter and M. Tonini, J. High Energy Phys. 02, 077 (2013); M. Tonini, Report No. DESY-THESIS-2014-038,
 PhD Thesis (Universit$\ddot{\textrm{a}}$t Hamburg, 2014).
\bibitem{SLH5}G. Marandella, C. Schappacher, and A. Strumia, Phys. Rev. D 72, 035014 (2005).
\bibitem{Zprime}The ATLAS Collaboration, J. High Energy Phys. 10, 182 (2017).
\bibitem{SSM}P. Langacker, Rev. Mod. Phys. 81, 1199 (2009).
\bibitem{MSTW}A. D. Martin, W. J. Stirling, R. S. Thorne, and G. Watt, Eur. Phys. J. C 63, 189 (2009); see
 also \url{http://mstwpdf.hepforge.org/}.
\bibitem{TParity}D. Dercks, G. Moortgat-Pick, J. Reuter, and S. Y. Shim, Report No. DESY-17-192, arXiv: 1801.06499.
\bibitem{LEP}The ALEPH, DELPHI, L3, and OPAL Collaborations (LEP Higgs Working Group), Report No. LHWG Note/2001-04, arXiv: hep-ex/0107030;
 G. Abbiendi {\em et al.} (ALEPH, DELPHI, L3, and OPAL Collaborations (the LEP Higgs Working Group)), Phys. Lett. B 565, 61 (2003);
 S. Schael {\em et al.} (ALEPH, DELPHI, L3, and OPAL Collaborations (the LEP Higgs Working Group)), Eur. Phys. J. C 47, 547 (2006).
\bibitem{exo}D. Curtin {\em et al.}, Phys. Rev. D 90, 075004 (2014).
\bibitem{h2ZZ}The ATLAS Collaboration, Report No. ATLAS-CONF-2017-058.
\bibitem{h2WW}The ATLAS Collaboration, Eur. Phys. J. C 78, 24 (2018).
\bibitem{hig12}A. Djouadi, Phys. Rep. 457, 1 (2008); Phys. Rep. 459, 1 (2008).
\bibitem{LHCW}The LHC Higgs Cross Section Working Group, Report No. CERN-2011-002, arXiv: 1101.0593; Reports No. CERN-2013-004
 and No. FERMILAB-CONF-13-667-T, arXiv: 1307.1347; Reports No. FERMILAB-FN-1025-T and No. CERN-2017-002-M, arXiv: 1610.07922.
\bibitem{EDMe}The ACME Collaboration, Science 343, 269 (2014).
\bibitem{EDMn}C. Baker {\em et al.}, Phys. Rev. Lett. 97, 131801 (2006); J. M. Pendlebury {\em et al.}, Phys. Rev. D 92, 092003 (2015).
\bibitem{BZ}S. M. Barr and A. Zee, Phys. Rev. Lett. 65, 21 (1990); Phys. Rev. Lett. 65, 2920 (1990, erratum).
\bibitem{BZ2}J. Brod, U. Haisch, and J. Zupan, J. High Energy Phys. 11, 180 (2013);
 T. Abe, J. Hisano, T. Kitahara, and K. Tobioka, J. High Energy Phys. 01, 106 (2014);
 K. Cheung, J. S. Lee, E. Senaha, and P.-Y. Tseng, J. High Energy Phys. 06, 149 (2014).
\bibitem{wein}S. Weinberg, Phys. Rev. Lett. 63, 2333 (1989); D. A. Dicus, Phys. Rev. D 41, 999 (1990);
 E. Braaten, C.-S. Li, and T.-C. Yuan, Phys. Rev. Lett. 64, 1709 (1990).
\bibitem{dHg}B. Graner, Y. Chen, E. G. Lindahl, and B. R. Heckel, Phys. Rev. Lett. 116, 161601 (2016); Phys. Rev. Lett. 119, 119901 (2017, erratum).
\bibitem{dnind} V. F. Dmitriev and R. A. Sen'kov, Phys. Rev. Lett. 91, 212303 (2003)
\bibitem{atomEDM}J. Engel, M. J. Ramsey-Musolf, and U. van Kolck, Prog. Part. Nucl. Phys. 71, 21 (2013); V. Cirigliano, W. Dekens, J. de Vries,
 and E. Mereghetti, Phys. Rev. D 94, 034031 (2016).
\bibitem{CEPC}The CEPC-SPPC Study Group, Reports No. IHEP-CEPC-DR-2015-01, No. IHEP-TH-2015-01, No. IHEP-EP-2015-01, and No. IHEP-AC-2015-01;
 \url{http://cepc.ihep.ac.cn/preCDR/volume.html}.
\bibitem{TLEP}M. Bicer {\em et al.} (TLEP Design Study Working Group), J. High Energy Phys. 01, 164 (2014).
\bibitem{sig}S. Heinemeyer and C. Schappacher, Eur. Phys J. C 76, 220 (2016).
\bibitem{rec}J. F. Gunion, T. Han, and R. Sobey, Phys. Lett. B 429, 79 (1998).
\bibitem{rec2}NLC ZDR Design Group and NLC Physics Working Group Collaborations, arXiv: hep-ex/9605011.
\bibitem{ptb}H. Li, arXiv: 1007.2999; PhD Thesis (Universit$\acute{\textrm{e}}$ de Paris-Sud, 2009),
 \url{http://hal.inria.fr/file/index/docid/430432/filename/Li.pdf}.
\bibitem{Zrare}J. Liu, \url{http://indico.ihep.ac.cn/event/6937/session/3/contribution/22/material/slides/0.pdf}; J. Liu, L.-T. Wang,
 X.-P. Wang, and W. Xue, arXiv: 1712.07237.
\bibitem{CMSFUT}The CMS Collaboration, Report No. CMS-PAS-FTR-13-024.
\bibitem{rare}Z. Liu, L.-T. Wang, and H. Zhang, Chin. Phys. C 41, 063102 (2017).
\bibitem{100T}N. Arkani-Hamed, T. Han, M. Mangano, and L.-T. Wang, Phys. Rep. 652, 1 (2016).

\end{thebibliography}
\end{document}